\documentclass[onecolumn,aps,12pt,longbibliography]{revtex4-1}
\usepackage{graphicx}
\usepackage{caption}
\usepackage{amsmath} 
\usepackage{float}
\usepackage{xcolor}

\renewcommand{\thefigure}{\textbf{\arabic{figure}}}

\usepackage{etoolbox}
\usepackage[section]{placeins}
\usepackage{lipsum}
\begin{document}

\def\bb{\begin{equation}}
\def\ee{\end{equation}}

%\title{Floquet Chern Insulators and Negative Resistance Catastrophe in Irradiated Graphene}
\title{Floquet Chern Insulators and Radiation-Induced Zero Resistance in Irradiated Graphene}

\author{Youngjae Kim and Kwon Park$^{\ast}$}
\affiliation{School of Physics, Korea Institute for Advanced Study, Seoul 02455, Korea}

\date{\today}

\begin{abstract}
Recent advances in optics and time-resolved techniques have facilitated the exploration of new states of matter under nonequilibrium conditions. 
Here, we predict that irradiated graphene can host two novel nonequilibrium steady states of matter with zero resistance when exposed to circularly polarized light: (i) Floquet Chern insulators and  (ii) a radiation-induced zero-resistance state with spontaneous formation of an inhomogeneous current distribution.
Specifically, we calculate nonequilibrium anomalous Hall and longitudinal conductivities to map the nonequilibrium phase diagram of irradiated graphene as a function of the driving frequency and the electric-field strength of circularly polarized light. 
As a result, Floquet Chern insulators are found to occur at high driving frequencies above the graphene band width. 
By contrast, at low driving frequencies below the graphene band width, the nonequilibrium anomalous Hall conductivity deviates from the expected quantized values, and the nonequilibrium longitudinal conductivity exhibits highly irregular behavior, including negative resistance. 
It is predicted that the thermodynamically unstable negative resistance will trigger a catastrophic breakdown, inducing a zero-resistance state with spontaneous formation of an inhomogeneous current distribution, similar to the radiation-induced zero-resistance state observed in quantum Hall systems. 
\end{abstract}

\maketitle

New states of matter are important because they can reveal novel physical mechanisms for organizing electrons, atoms, and molecules into unexpected collective states.
One particular example of such unexpected collective states is the zero-resistance state. 
Conventionally, zero resistance can be induced via two physical mechanisms: (i) Cooper pairing in superconductors and (ii) topological edge protection in quantum Hall states and Chern insulators. 
There is, however, a lesser-known third mechanism for the radiation-induced zero-resistance (RIZR) state in quantum Hall systems exposed to microwave~\cite{Mani1,Mani2,Zudov1,Shi1,Durst1,Park1,Durst2,Andreev1}, referred to as mechanism (iii).

The quantum Hall RIZR state is an emergent nonequilibrium steady state of matter that cannot be understood in terms of equilibrium physics. 
Several key ingredients are necessary to understand the physical mechanism behind the quantum Hall RIZR state.
First, the time-periodic driving field generates an infinite set of quasienergy eigenstates known as Floquet eigenstates~\cite{Oka2,Usaj1,Sentef1,Oka1,Oka1,Kim1,Rudner1,McIver1,Alberto1,Shan1,Park2,Bao1,Boschini1,Tyulnev1}, which have been shown to form quasienergy bands in solids through spectroscopic measurements~\cite{Wang1,Aeschlimann1,Zhou1,Ito1,Merboldt1,Choi1}.
Second, Floquet eigenstates are populated via a novel nonequilibrium distribution function that differs from the usual Fermi-Dirac distribution function, which can lead to population inversion under certain conditions~\cite{Shi1,Durst1,Park1,Durst2}. 
Population inversion is a phenomenon in which higher energy levels are more populated than lower energy levels.
Third, population inversion can cause negative resistance.
Fourth, the thermodynamically unstable negative resistance triggers a catastrophic breakdown, called the negative resistance catastrophe (NRC), which induces a zero-resistance state with spontaneous formation of an inhomogeneous current distribution~\cite{Andreev1,Auerbach1,Finkler1,Dmitriev1,Dmitriev2,Dorozhkin2,Wang2}.

Graphene offers another exciting possibility of achieving zero resistance through radiation-induced topological matter, known as Floquet Chern insulators~\cite{Oka2,Usaj1,Sentef1,Oka1,Kim1,McIver1}.
Floquet Chern insulators are a type of Chern insulator where Floquet eigenstates form topologically nontrivial Chern bands.   
Zero resistance can be induced in irradiated graphene via topological protection of edge states in Floquet Chern bands, referred to as mechanism (iv). 
This work predicts that a new, nested form of the RIZR state—the NRC-induced RIZR state arising from Floquet Chern insulators—can occur in irradiated graphene by combining mechanisms (iii) and (iv), which is then referred to as mechanism (v).

Specifically, we investigate the nonequilibrium phase diagram of irradiated graphene as a function of the driving frequency and the electric-field strength of circularly polarized light.
To this end, we calculate nonequilibrium anomalous Hall and longitudinal conductivities of irradiated graphene using the nonequilibrium Green's function method. 
A key step in our implementation of the nonequilibrium Green's function method is the Floquet eigenstate thermalization (FET) hypothesis, which involves populating the Floquet eigenstates using the same fundamental thermalization scheme employed for the quantum Hall RIZR state.

As a result, it is found that in the regime where the driving frequency is higher than the graphene band width, i.e., $\Omega > {\cal W}$, the nonequilibrium anomalous Hall conductivity reproduces well-quantized Chern numbers of the Floquet bands, consistent with the na\"{i}ve expectation that Floquet eigenstates behave as if they were the usual energy eigenstates in equilibrium.
Meanwhile, the nonequilibrium longitudinal conductivity is almost zero everywhere, except at the boundaries where Chern numbers change. 
The quantized anomalous Hall conductivity and zero longitudinal conductivity together indicate the emergence of Floquet Chern insulators in this regime.
By contrast, in the regime where the driving frequency is lower than the graphene band width, i.e., $\Omega < {\cal W}$, the nonequilibrium anomalous Hall conductivity deviates from the expected quantized values, and the nonequilibrium longitudinal conductivity exhibits highly irregular behavior, including negative resistance. 
It is predicted that the thermodynamically unstable negative resistance will trigger a catastrophic breakdown in Floquet Chern insulators, inducing a zero-resistance state with spontaneous formation of an inhomogeneous current distribution, similar to the quantum Hall RIZR state.

\section*{Results}

\subsection*{Floquet Hamiltonian and nonequilibrium Green's functions}

\begin{figure*}[ht]
\includegraphics[width=1\textwidth]{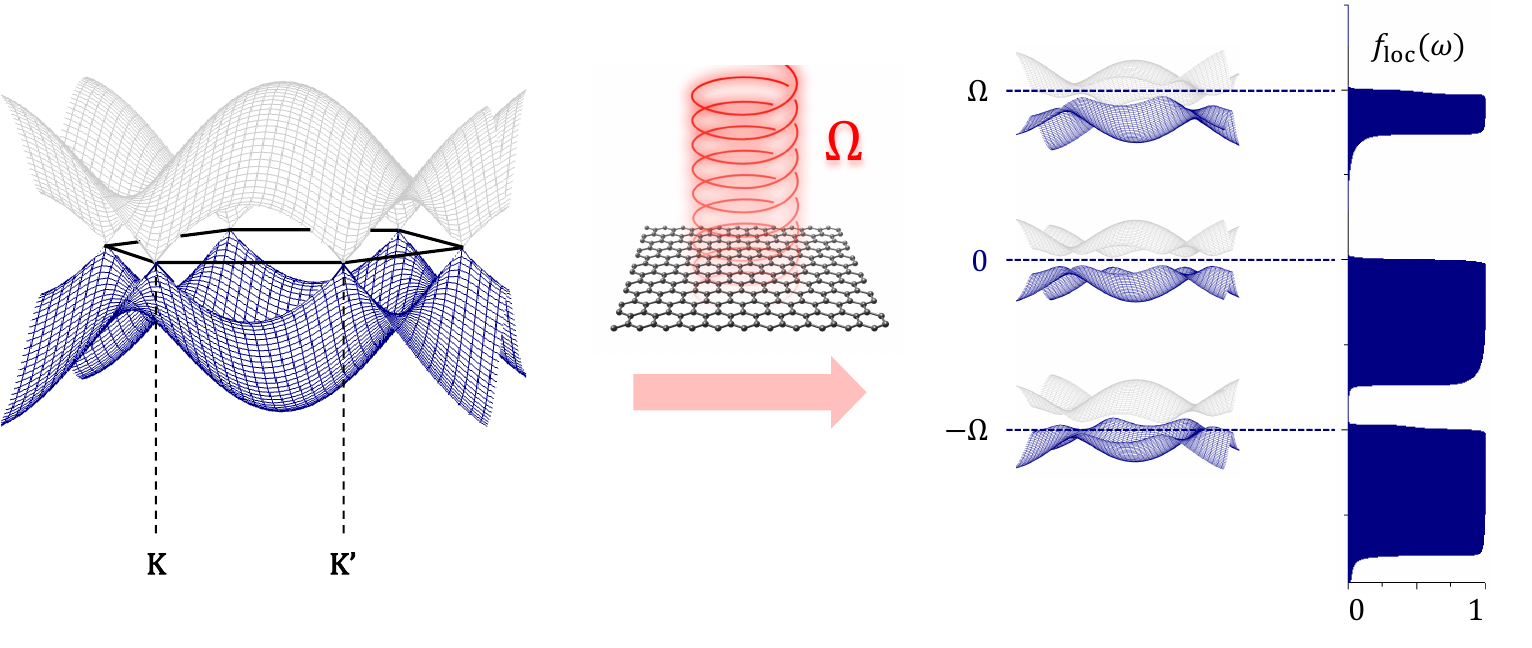}
\caption{
\textbf{Emergence of a Floquet Chern insulator in irradiated Graphene.}
The left part of the figure shows the energy band structure of graphene at half-filling before irradiation, where the blue and gray energy bands represent occupied and unoccupied states, respectively. 
Two Dirac cones are located at the corners of the first Brillouin zone, labeled as ${\bf K}$ and ${\bf K}^\prime$.
The middle part of the figure shows that graphene is irradiated with circularly polarized light at the driving frequency $\Omega$. 
In this figure, $\Omega = 8\tau$, where $\tau$ is the hopping parameter between nearest-neighbor carbon atoms.
The right part of the figure shows that, after irradiation, an infinite set of well-separated Floquet copies of two quasienergy sub-bands forms, with the overall Floquet energy gap of $\Omega$. 
The two quasienergy sub-bands have the same energetic and topological structure as the energy bands in the honeycomb Haldane model, with a small energy gap opening at ${\bf K}$ and ${\bf K}^\prime$.
The local distribution function, $f_{\rm loc}(\omega)$, indicates the average population of Floquet eigenstates across momenta as a function
of quasienergy.
}
\label{FIG1}
\end{figure*}

We start with the Hamiltonian describing graphene before irradiation:
%%%%%%%%%%%%%
\begin{equation}
\hat{H}_0({\mathbf{k}}) = 
\begin{pmatrix}
0 & \Delta_{\mathbf{k}} \\
\Delta_{\mathbf{k}}^* & 0
\end{pmatrix},
\label{eq:H_0}
\end{equation}
%%%%%%%%%%%%%
where $\Delta_{\mathbf{k}} = -\tau \left(e^{i\mathbf{k} \cdot \mathbf{a}_{1}} + e^{i\mathbf{k} \cdot \mathbf{a}_{2}} + e^{i\mathbf{k} \cdot \mathbf{a}_{3}} \right)$ with $\mathbf{a}_1 = l(0,1)$, $\mathbf{a}_2 = -\frac{l}{2}(\sqrt{3},1)$, and $\mathbf{a}_3 = \frac{l}{2}(\sqrt{3},-1)$ with $\tau$ being the hopping parameter and $l$ being the distance between nearest-neighbor carbon atoms.
The honeycomb lattice structure provides two Dirac cones at the corners of the first Brillouin zone, $\mathbf{K}$ and $\mathbf{K^\prime}$, in reciprocal space. 
The energy band structure is given as $\pm |\Delta_{\mathbf{k}}|$ with a band width of ${\cal W} = 6\tau$, as shown on the left side of Fig.~\ref{FIG1}.

The Hamiltonian describing irradiated graphene exposed to circularly polarized light can be obtained from Eq.~\eqref{eq:H_0} after applying the Peierls substitution, $\mathbf{k} \rightarrow \mathbf{k} - e\mathbf{A}(t)/\hbar c$, with $\mathbf{A}(t) = A(\cos{\Omega t}, \sin{\Omega t})$:
%%%%%%%%%%%%%%%%%%%%%%%%%%%%%%%%%%%%%
\begin{equation}
\hat{H}_{\mathbf{k}}(t) = \hat{H}_0({{\bf k} \rightarrow {\bf k}-e{\bf A}(t)/\hbar c}),
\label{eq:H_t}
\end{equation}
%%%%%%%%%%%%%%%%%%%%%%%%%%%%%%%%%%%%%
where $A$ is related to the electric-field strength $E_0$ via $A=c E_0/\Omega$.
From this point forward, we set $\tau$, $l$, $e$, $\hbar$ and $c$ to unity for simplicity unless stated otherwise.

According to Floquet theory, solving the time-dependent Schrödinger equation for a time-periodic Hamiltonian amounts to diagonalizing an effective Hamiltonian known as the Floquet Hamiltonian~\cite{Oka1,Kim1}:
%%%%%%%%%%%%%%%%%%%%%%%%%%%%%%%%%%%%%%%%%%%%%%%%%%%%
\begin{equation}
[\hat{H}^{\rm F}_{\mathbf{k}}]_{nm} = [\hat{H}_{\mathbf{k}}]_{nm} + n \Omega \delta_{nm} \mathbf{1} ,    
\label{eq:H^F}
\end{equation}
%%%%%%%%%%%%%%%%%%%%%%%%%%%%%%%%%%%%%%%%%%%%%%%%%%%%
where $[\hat{H}_{\mathbf{k}}]_{nm}$ denotes a Fourier component of the $2 \times 2$ time-periodic Hamiltonian: 
%%%%%%%%%%%%%%%%%%%%%%%%%%%%%%%%%%%%%%%%%%%%%%%
\begin{equation}
[\hat{H}_{\mathbf{k}}]_{nm} = \frac{1}{T} \int_0^T dt \hat{H}_{\mathbf{k}}(t) e^{i(n-m)\Omega t} ,
\end{equation}
%%%%%%%%%%%%%%%%%%%%%%%%%%%%%%%%%%%%%%%%%%%%%%%
where $n$ and $m$ are the Floquet indices.
It is important to note that the second term on the right side of Eq.~\eqref{eq:H^F} represents the linear potential term caused by a constant effective electric field with a strength of $\Omega$ applied in the Floquet index space.
Formally, the Floquet Hamiltonian in Eq.~\eqref{eq:H^F} is exactly equivalent to the Stark Hamiltonian with an effective electric field applied in the Floquet index space~\cite{Kim1}.
In this context, Floquet eigenstates, which diagonalize the Floquet Hamiltonian, can be understood as the Wannier-Stark ladder (WSL) eigenstates resulting from the Stark effect in the Floquet index space.

Next, we need to account for the effects of impurity scattering to calculate the nonequilibrium DC electric conductivities. 
To this end, it is useful to employ the nonequilibrium Green's function method, which starts with the noninteracting Green's functions in the absence of impurity scattering:
%%%%%%%%%%%%%%%%%%%%%%%%%%%%%%%%%%%%%%%%%%%%%%%%%%%%%%%%%%%%%%
\begin{align}
G^{r,<}_{0}(\mathbf{k},\omega) = \mathcal{U}_{\mathbf{k}} {\cal G}_{\rm F}^{r,<}(\mathbf{k},\omega) \mathcal{U}^{\dagger}_{\mathbf{k}} ,
\label{eq:G_0}
\end{align}
%%%%%%%%%%%%%%%%%%%%%%%%%%%%%%%%%%%%%%%%%%%%%%%%%%%%%%%%%%%%%%
where $G^r_0(\mathbf{k},\omega)$ and $G^<_0(\mathbf{k},\omega)$ denote the noninteracting retarded and lesser Green's functions of the Bloch state, respectively. 
%which are connected to those of the Floquet eigenstate, $g^r(\mathbf{k},\omega)$ and $g^<(\mathbf{k},\omega)$, via the unitary transformation.
Equation~\eqref{eq:G_0} indicates that the noninteracting Green's functions of the Bloch state, $G^r_0(\mathbf{k},\omega)$ and $G^<_0(\mathbf{k},\omega)$, are connected to those of the Floquet eigenstate, ${\cal G}_{\rm F}^r(\mathbf{k},\omega)$ and ${\cal G}_{\rm F}^<(\mathbf{k},\omega)$, through a unitary transformation.
Specifically, $\mathcal{U}_{\mathbf{k}}$ is the unitary operator diagonalizing the Floquet Hamiltonian: 
%%%%%%%%%%%%%%%%%%%%%%%%%%%%%%%%%%%%%%%%%%%%%%%%%%%%%%%%%%%%%%%
\begin{equation}
\mathcal{U}^{\dagger}_{\mathbf{k}} H^{\rm F}_{\mathbf{k}}\mathcal{U}_{\mathbf{k}} = {\rm diag}(\epsilon_{\mu,n,\mathbf{k}}),
\label{eq:U}
\end{equation}
%%%%%%%%%%%%%%%%%%%%%%%%%%%%%%%%%%%%%%%%%%%%%%%%%%%%%%%%%%%%%%% 
where $H^{\rm F}_{\bf k}$ represents the Floquet Hamiltonian expressed as a matrix with combined indices for both sublattices and Floquet components, and $\epsilon_{\mu,n,\mathbf{k}}$ denotes the $\mu$-th quasienergy sub-band of the $n$-th Floquet level, i.e., $\epsilon_{\mu,n,\mathbf{k}}=\epsilon_{\mu,\mathbf{k}}+n\Omega$.
In the case of irradiated graphene, $\mu$ can have two values: $+$ and $-$, while $n$ can take any integer value.

The retarded Green's function of the Floquet eigenstate, ${\cal G}_{\rm F}^r(\mathbf{k},\omega)$, contains information about quasienergy eigenvalues, expressed as follows~\cite{Tsuji1}:
%%%%%%%%%%%%%%%%%%%%%%%%%%%%%%%%%%%%%%%%%%%%%%%%%%%%%%%%%%%%%%%
\begin{align}
[{\cal G}_{\rm F}^r(\mathbf{k},\omega)]_{\mu n,\nu m}= \frac{1}{\omega-\epsilon_{\mu,\mathbf{k}}-n\Omega+i\delta}\delta_{\mu\nu}\delta_{nm} ,
\label{eq:cal_G^r}
\end{align}
%%%%%%%%%%%%%%%%%%%%%%%%%%%%%%%%%%%%%%%%%%%%%%%%%%%%%%%%%%%%%%%
where $\omega$ is restricted within the first Floquet Brillouin zone, i.e., $-\Omega/2 < \omega \le \Omega/2$.
Meanwhile, the lesser Green's function of the Floquet eigenstate, ${\cal G}_{\rm F}^<(\mathbf{k},\omega)$, contains information about nonequilibrium thermalization.
To obtain ${\cal G}_{\rm F}^<(\mathbf{k},\omega)$, we employ the FET hypothesis, which is the same fundamental thermalization scheme used for the quantum Hall RIZR state, with a crucial extension to account for dispersive quasienergy bands.
Specifically, ${\cal G}_{\rm F}^<(\mathbf{k},\omega)$ can be expressed in the standard form required by the fluctuation-dissipation theorem, except for the chemical potential shift individually adjusted for each Floquet eigenstate:
%%%%%%%%%%%%%%%%%%%%%%%%%%%%%%%%%%%%%%%%%%%%%%%%%%%%%%%%%%%%%%%%%%%%
\begin{align}
{\cal G}_{\rm F}^<(\mathbf{k},\omega)= \left( {\cal G}_{\rm F}^{a}(\mathbf{k},\omega)-{\cal G}_{\rm F}^{r}(\mathbf{k},\omega) \right) \mathcal{F}_{\rm FD}(\mathbf{k},\omega) ,
\label{eq:cal_G^<}
\end{align}
%%%%%%%%%%%%%%%%%%%%%%%%%%%%%%%%%%%%%%%%%%%%%%%%%%%%%%%%%%%%%%%%%%%%
where
${\cal G}_{\rm F}^{a}(\mathbf{k},\omega)=[{\cal G}_{\rm F}^{r}(\mathbf{k},\omega)]^\dagger$ is the advanced Green's function of the Floquet eigenstate, and
%%%%%%%%%%%%%%%%%%%%%%%%%%%%%%%%%%%%%%%%%%%%%%%%%%%%%%%%%%%%%
\begin{align}
[\mathcal{F}_{\rm FD}(\mathbf{k},\omega)]_{\mu n,\nu m}=f_{\rm FD} \left(
\omega-\langle \mathcal{N} \rangle_{{\mu,n,\mathbf{k}}} \Omega \right)
\delta_{\mu\nu}\delta_{nm} ,
\label{eq:cal_F}
\end{align}
%%%%%%%%%%%%%%%%%%%%%%%%%%%%%%%%%%%%%%%%%%%%%%%%%%%%%%%%%%%%%
where $f_{\rm FD}(\epsilon)$ is the usual Fermi-Dirac distribution function with the temperature set to zero in this work. 
It is important to note that the chemical potential shift is individually adjusted for each Floquet eigenstate by the expectation value of the Floquet index operator: $\langle \mathcal{N} \rangle_{{\mu,n,\mathbf{k}}}=\langle \phi_{\mu,n,\mathbf{k}} | \mathcal{N} | \phi_{\mu,n,\mathbf{k}} \rangle$, where $[{\mathcal{N}}]_{\mu n,\nu m}=n\delta_{\mu\nu}\delta_{nm}$ is the Floquet index operator, and $|\phi_{\mu,n,\mathbf{k}}\rangle$ is the Floquet eigenstate with quasienergy eigenvalue $\epsilon_{\mu,n,\mathbf{k}}$.
In the case of irradiated quantum Hall systems, where quasienergy bands are completely dispersionless, $\langle \mathcal{N} \rangle_{{\mu,n,\mathbf{k}}}$ simplifies to $n$, reproducing exactly the same thermalization scheme used for the quantum RIZR state~\cite{Shi1, Durst1, Park1}.

The FET hypothesis can be understood by drawing an analogy to the WSL thermalization scheme, which can be obtained by solving the Lindblad equation describing the dynamics of electrons in the lattice, exposed to a constant electric field while coupled with a bosonic thermal bath~\cite{Lee1, Hwang1}.  
The essence of the WSL thermalization scheme is to adjust the chemical potential shift to ensure a uniform charge distribution in real space. 
Likewise, the FET hypothesis guarantees a uniform charge distribution in the Floquet index space.

After constructing $G_0^{r,<}(\mathbf{k},\omega)$ via Eqs.~\eqref{eq:G_0}, \eqref{eq:U}, \eqref{eq:cal_G^r}, \eqref{eq:cal_G^<}, and \eqref{eq:cal_F}, one can plug them into the Keldysh-Dyson equations to self-consistently determine the fully interacting retarded and lesser Green's functions, $G^{r,<}(\mathbf{k},\omega)$:
%%%%%%%%%%%%%%%%%%%%%%%%%%%%%%%%%%%%%%%%%%%%%%%%%%%%%%%%%%%%%%%%%%%
\begin{align}
G^r &= \left( [G_0^r]^{-1} - \Sigma^r \right)^{-1} , \\
G^< &= G^{r} \left( [G_0^r]^{-1} G_0^< [G_0^a]^{-1} + \Sigma^{<} \right) G^{a} ,
\label{eq:Keldysh-Dyson}
\end{align}
%%%%%%%%%%%%%%%%%%%%%%%%%%%%%%%%%%%%%%%%%%%%%%%%%%%%%%%%%%%%%%%%%%%
where $\Sigma^{r,<}$ represents the retarded and lesser self-energies.
For brevity, the arguments of the Green's functions, $\mathbf{k}$ and $\omega$, have been omitted above.
The self-energies are calculated using the self-consistent Born approximation (SCBA) of delta-function impurity scattering~\cite{Lee1,Hwang1}: 
%%%%%%%%%%%%%%%%%%%%%%%%%%%%%%%%%%%%%%%%%%%%%%%%%%%%%%%%%%%%%%%%%%%%%%%%%%
\begin{equation}
[\Sigma^{r,<}(\omega)]_{in,jm} = v_\mathrm{imp}^2 \int \frac{d^{2}{\mathbf{k}}}{(2\pi)^2} [G^{r,<}(\mathbf{k},\omega)]_{in,im} \delta_{ij} ,
\label{eq:SCBA}
\end{equation}
%%%%%%%%%%%%%%%%%%%%%%%%%%%%%%%%%%%%%%%%%%%%%%%%%%%%%%%%%%%%%%%%%%%%%%%%%%
where $i$ and $j$ denote the sublattice indices, and $v_\mathrm{imp}$ represents the strength of impurity scattering. 
Here, we set $v_\mathrm{imp}=0.6$ unless stated otherwise.
Note that self-energies are diagonal in the sublattice index because delta-function impurity scattering does not mix different sublattices.

The fully interacting retarded and lesser Green's functions can offer physical insights into the resulting nonequilibrium steady state of matter through the local density of states (LDOS), defined as:
%%%%%%%%%%%%%%%%%%%%%%%%%%%%%%%%%%%%%%%%%%%%%%%%%%%%%%%%%%%%%%%%%%%
\begin{align}
\rho_{\rm loc}(\omega+n\Omega) = -\frac{1}{\pi} \int \frac{d^{2}{\mathbf{k}}}{(2\pi)^2}  \sum_i \mathrm{Im} [G^r(\mathbf{k},\omega)]_{in,in} , 
\end{align}
%%%%%%%%%%%%%%%%%%%%%%%%%%%%%%%%%%%%%%%%%%%%%%%%%%%%%%%%%%%%%%%%%%%
and the local occupation number (LON), defined as:
%%%%%%%%%%%%%%%%%%%%%%%%%%%%%%%%%%%%%%%%%%%%%%%%%%%%%%%%%%%%%%%%%%%
\begin{align}
N_{\rm loc}(\omega+n\Omega) = \frac{1}{2\pi} \int \frac{d^{2}{\mathbf{k}}}{(2\pi)^2}  \sum_i \mathrm{Im} [G^<(\mathbf{k},\omega)]_{in,in} . 
\end{align}
%%%%%%%%%%%%%%%%%%%%%%%%%%%%%%%%%%%%%%%%%%%%%%%%%%%%%%%%%%%%%%%%%%%
The ratio of LON to LDOS determines the local distribution function, $f_{\rm loc}(\omega) = N_{\rm loc}(\omega)/\rho_{\rm loc}(\omega)$, which indicates the average population of Floquet eigenstates across different momenta as a function of quasienergy.

Figure~\ref{FIG1} illustrates the formation of a Floquet Chern insulator in irradiated graphene exposed to circularly polarized light with a driving frequency sufficiently higher than the graphene band width, $\Omega=8 \tau$. 
In this regime, the energy band of graphene evolves into an infinite set of well-separated Floquet copies of two quasienergy sub-bands, with a small energy gap opening at the Dirac points, which are precisely equivalent to the two energy bands of the honeycomb Haldane model~\cite{Kim1,Haldane1,Oka2,Mikami1}.

At first glance, the local distribution function looks quite different from the Fermi-Dirac distribution function, $f_{\rm FD}(\omega)$, in equilibrium.
However, a close inspection reveals that the lower quasienergy sub-band of each Floquet copy is fully occupied, as if it were under $f_{\rm FD}(\omega)$. 
This indicates that the nonequilibrium anomalous Hall conductivity should exhibit well-quantized values determined by the Chern number of the fully occupied, lower quasienergy sub-band.
This is indeed confirmed by the actual calculation of nonequilibrium DC electric conductivities, as shown below.

\subsection*{Nonequilibrium DC electric conductivities of irradiated graphene}

The nonequilibrium DC electric conductivities can be calculated by inserting the fully interacting Green's functions into the nonequilibrium Kubo formula:
%%%%%%%%%%%%%%%%%%%%%%%%%%%%%%%%%%%%%%%%%%%%%%%%%%%%%%%%%%%%%%%%%%%%%%%%%%%%
\begin{equation}
\sigma_{\alpha \beta}^{\rm full} =  \int \frac{d^{2}\mathbf{k}}{(2\pi)^2} \int \frac{d\omega}{2\pi} 
\mathrm{Tr} 
\left( 
J^{\alpha}_{\mathbf{k}} \frac{\partial G^{r}({\mathbf{k}},\omega)}{\partial \omega} J^{\beta}_{\mathbf{k}} G^{<}(\mathbf{k},\omega) 
-J^{\alpha}_{\mathbf{k}} G^{<}(\mathbf{k},\omega) {J}^{\beta}_{\mathbf{k}}\frac{\partial G^{a}(\mathbf{k},\omega)}{\partial \omega}
\right),
\label{eq:noneq_Kubo}
\end{equation}
%%%%%%%%%%%%%%%%%%%%%%%%%%%%%%%%%%%%%%%%%%%%%%%%%%%%%%%%%%%%%%%%%%%%%%%%%%%%
where $[{J}^{\alpha}_{\mathbf{k}}]_{in,jm} = -\frac{\partial}{\partial k_{\alpha}} [\hat{H}_{{\mathbf{k}},ij}]_{nm}$ is the current operator at momentum ${\bf k}$ in the Floquet representation.
Here, the frequency integration is performed within the first Floquet Brillouin zone, i.e., $-\Omega/2<\omega\leq\Omega/2$.
%(As mentioned previously, the charge of an electron, $e$, is set to unity in this work.)
Here, ${\alpha}$ and ${\beta}$ denote spatial directions, either $x$ or $y$.
Note that, in Eq.~\eqref{eq:noneq_Kubo}, we only consider self-energy corrections and neglect vertex corrections caused by impurity scattering, assuming that impurity scattering is sufficiently weak.
Refer to Appendix~\ref{appen:noneq_Kubo} for details on the derivation of Eq.~\eqref{eq:noneq_Kubo}.

\begin{figure*}[ht]
\includegraphics[width=1.0\textwidth]{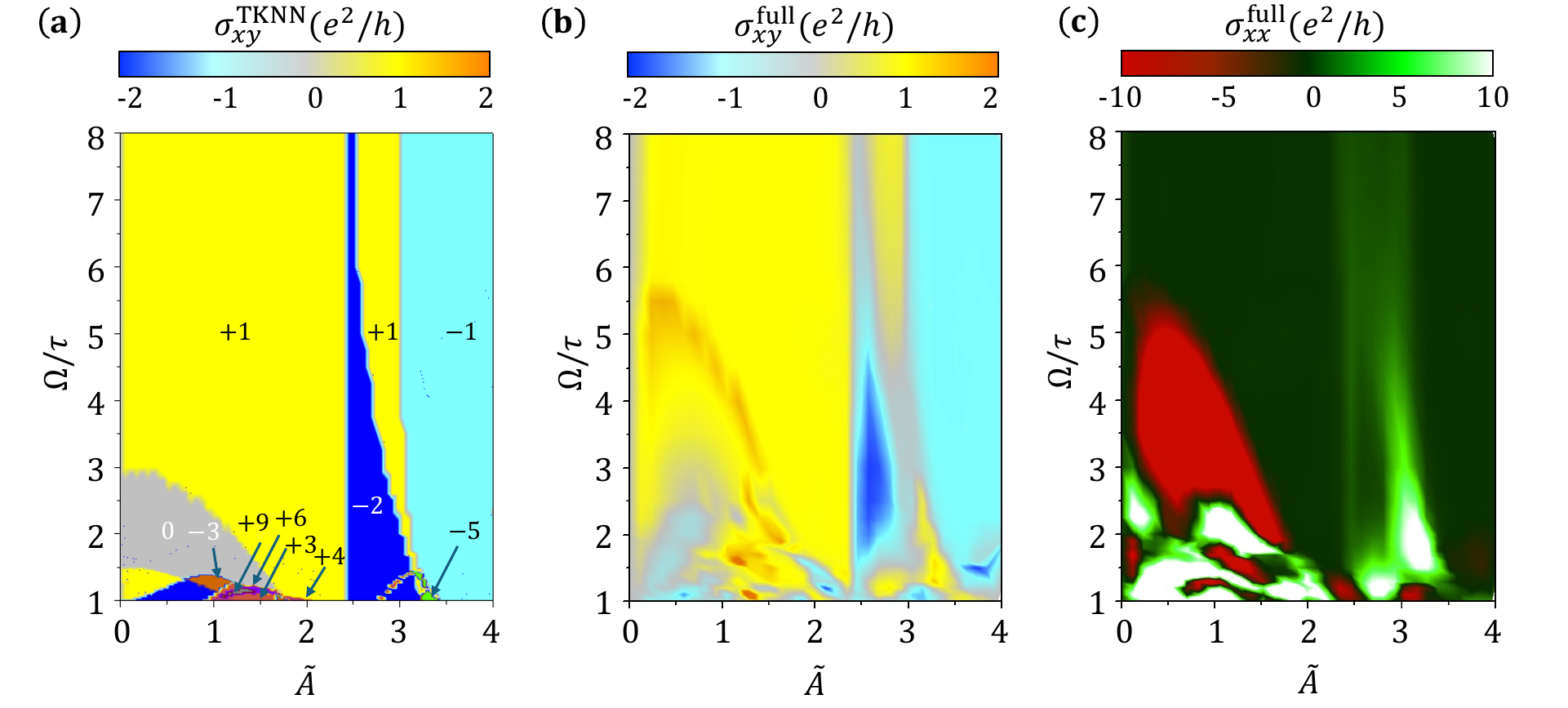}
\caption{
\textbf{Nonequilibrium DC electric conductivities of irradiated graphene.}
({\bf a}) Anomalous Hall conductivity calculated from the TKNN formula, $\sigma^{\rm TKNN}_{xy}$, as a function of the driving frequency, $\Omega/\tau$, and the normalized electric-field strength, $\tilde{A}=eA l/{\hbar c}=eE_0 l/\hbar \Omega$, where $E_0$ is the electric-field strength, and $l$ is the distance between nearest-neighbor carbon atoms.
({\bf b}) Nonequilibrium anomalous Hall conductivity calculated from the full nonequilibrium Kubo formula, $\sigma^{\rm full}_{xy}$.
({\bf c}) Nonequilibrium longitudinal conductivity calculated from the full nonequilibrium Kubo formula, $\sigma^{\rm full}_{xx}$.
The color bars at the top represent the conductivity values in units of $e^2/h$.
}
\label{FIG2}
\end{figure*}

Figure~\ref{FIG2} shows the results of both nonequilibrium anomalous Hall and longitudinal conductivities, $\sigma^{\rm full}_{xy}$ and $\sigma^{\rm full}_{xx}$, obtained from Eq.~\eqref{eq:noneq_Kubo} as a function of the driving frequency, $\Omega/\tau$, and the normalized electric-field strength, $\tilde{A}=eA l/{\hbar c}=eE_0 l/\hbar \Omega$, of the circularly polarized light. 
Several points need to be mentioned in order.

First, 
%$\sigma^{\rm full}_{xy}$ can identify the regimes where Floquet Chern insulators appear.
$\sigma^{\rm full}_{xy}$ is not always quantized, contrary to what the TKNN formula indicates in equilibrium:
%%%%%%%%%%%%%%%%%%%%%%%%%%%%%%%%%%%%%%%%%%%%%%%%%%%%%%%%%%%%%%%%%%%%%%%%%%
\begin{align}
\sigma^{\rm TKNN}_{xy} = \int \frac{d^{2}{\bf k}}{(2\pi)^2} \sum_{\mu,n} {\cal B}_{\mu,n,{\bf k}} f_{\rm FD}(\epsilon_{\mu,n,{\bf k}}),
\label{eq:TKNN}
\end{align}
%%%%%%%%%%%%%%%%%%%%%%%%%%%%%%%%%%%%%%%%%%%%%%%%%%%%%%%%%%%%%%%%%%%%%%%%%%
where ${\cal B}_{\mu,n,{\bf k}}$ is the Berry curvature of the Floquet eigenstate with the quasienergy $\epsilon_{\mu,n,{\bf k}}$.
%$\epsilon_{\mu,n,{\bf k}}=\epsilon_{\mu,{\bf k}}+n\Omega$.
As shown in Fig.~\ref{FIG2}~({\bf a}), $\sigma^{\rm TKNN}_{xy}$ is precisely quantized in units of $e^2/h$, representing the Chern number of the highest occupied quasienergy band, which has simple values of $\pm 1$ in the limit of high $\Omega/\tau$, but develops complex structures as $\Omega/\tau$ decreases~\cite{Mikami1}.
Figure~\ref{FIG2}~({\bf b}) shows that $\sigma^{\rm full}_{xy}$ only matches $\sigma^{\rm TKNN}_{xy}$ in regimes where either $\Omega/\tau$ or $\tilde{A}$ is sufficiently high.
At the same time, Figure~\ref{FIG2}~({\bf c}) shows that $\sigma^{\rm full}_{xx}$ is zero whenever $\sigma^{\rm full}_{xy}$ is precisely quantized.
This indicates that Floquet Chern insulators appear in this regime.
As shown in Fig.~\ref{FIG2}~({\bf c}), $\sigma^{\rm full}_{xx}$ becomes positive at the phase boundaries where $\sigma^{\rm full}_{xy}$ switches between two different quantized values, indicating metallic behavior at topological phase transitions.

Second, $\sigma^{\rm full}_{xy}$ significantly deviates from $\sigma^{\rm TKNN}_{xy}$ in the regime of low $\Omega/\tau$ and $\tilde{A}$.
Most notably, as soon as $\Omega/\tau$ falls below the graphene band width ${\cal W}/\tau=6$, at which Floquet eigenstates begin to overlap with each other, $\sigma^{\rm full}_{xy}$ exhibits an overshoot from the quantized value, as shown by the orange curve on the yellow background in Fig.~\ref{FIG2}~({\bf b}). 
This overshoot curve indicates the emergence of a new nonequilibrium phase of matter, which is not a Floquet Chern insulator. 
The nature of this emergent nonequilibrium state of matter will be discussed in more detail below.
When $\Omega/\tau$ falls further below half the graphene  band width, $\sigma^{\rm full}_{xy}$ exhibits highly irregular behaviors, being almost completely detached from $\sigma^{\rm TKNN}_{xy}$, especially at small $\tilde{A}$.

Third, to understand why $\sigma^{\rm full}_{xy}$ is not always quantized, it is helpful to simplify the nonequilibrium Kubo formula for anomalous Hall conductivity in Eq.~\eqref{eq:noneq_Kubo} in the impurity-free limit as follows:
%%%%%%%%%%%%%%%%%%%%%%%%%%%%%%%%%%%%%%%%%%%%%%%%%%%%%%%
\begin{align}
\sigma^{\rm free}_{xy} =  \int \frac{d^{2}\mathbf{k}}{(2\pi)^2} \sum_{\mu,n} \mathcal{B}_{\mu,n,\mathbf{k}} 
f_{\rm FD}\left(\epsilon_{\mu,n,\mathbf{k}} - \langle \mathcal{N} \rangle_{\mu,n,\mathbf{k}} \Omega\right).
\label{eq:sigma_xy_free}
\end{align}
%%%%%%%%%%%%%%%%%%%%%%%%%%%%%%%%%%%%%%%%%%%%%%%%%%%%%%%
See Appendix~\ref{appen:sigma_xy_free} for a detailed derivation of Eq.~\eqref{eq:sigma_xy_free}.
As one can see, the key difference between $\sigma^{\rm free}_{xy}$ and $\sigma^{\rm TKNN}_{xy}$ is that the chemical potential is appropriately shifted in Eq.~\eqref{eq:sigma_xy_free} according to the FET hypothesis, whereas no such shift exists in Eq.~\eqref{eq:TKNN}.
Note that the chemical potential shift in Eq.~\eqref{eq:sigma_xy_free} is adjusted for each Floquet state by the expectation value of the Floquet index operator, $\langle \mathcal{N} \rangle_{\mu,n,\mathbf{k}}$, which depends on the quasienergy band indices, $\mu$ and $n$, and momentum, ${\bf k}$.  
This means that, under certain conditions, the higher quasienergy band can be partially filled, while the lower quasienergy band is partially empty. 
This phenomenon is called population inversion.

\begin{figure*}[ht]
\includegraphics[width=1.0\textwidth]{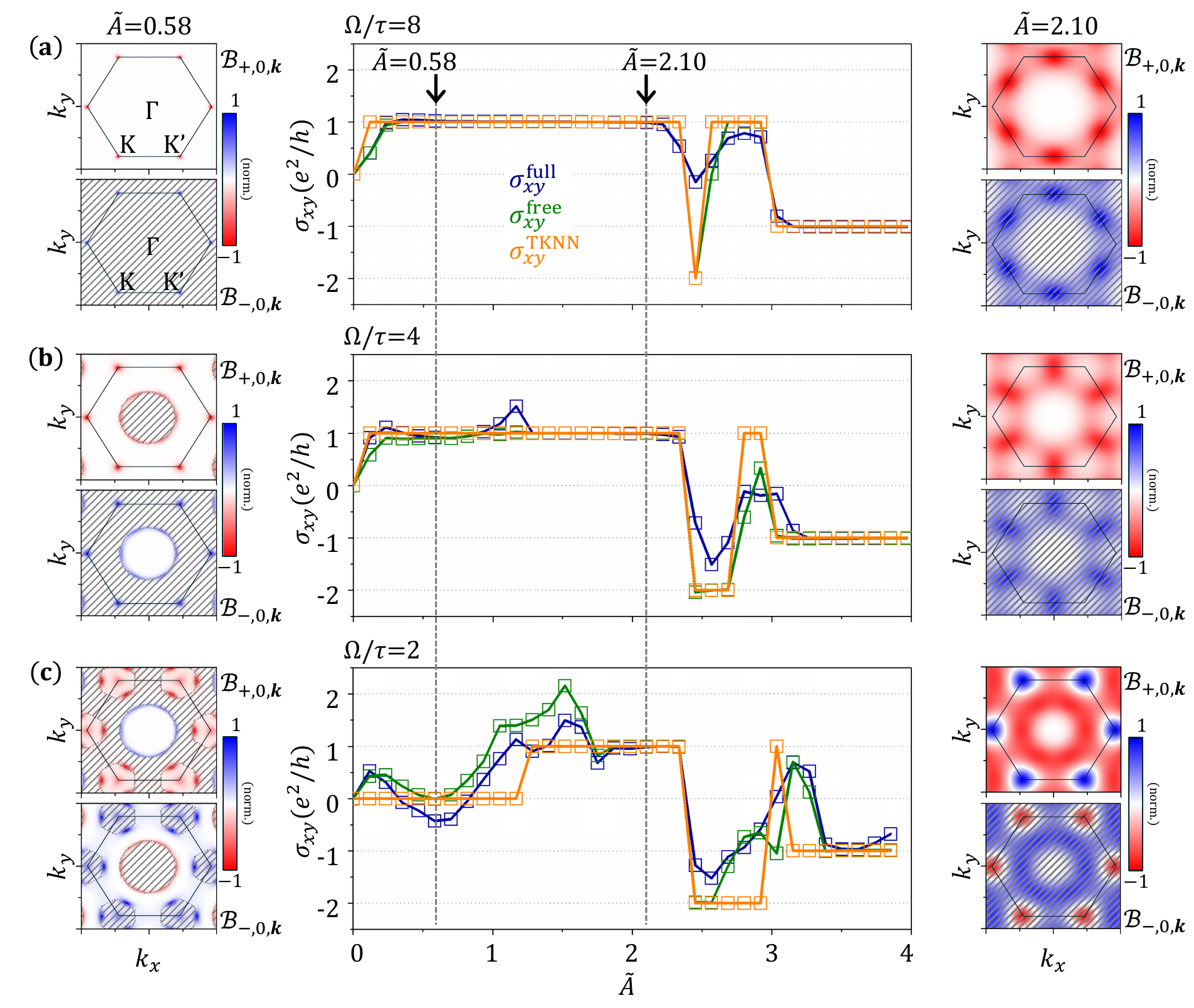}
\caption{
\textbf{Nonequilibrium anomalous Hall conductivity of irradiated graphene.}
The nonequilibrium anomalous Hall conductivity calculated from the fully interacting Green's functions, $\sigma_{xy}^{\rm full}$, is compared with that in the impurity-free limit, $\sigma_{xy}^{\rm free}$, and with that calculated using the TKNN formula, $\sigma_{xy}^{\rm TKNN}$. 
The comparison is conducted at three driving frequencies of ({\bf a}) $\Omega/\tau=8$, ({\bf b}) $\Omega/\tau=4$, and ({\bf c}) $\Omega/\tau=2$.
As shown in the comparison, $\sigma_{xy}^{\rm free}$ closely matches $\sigma_{xy}^{\rm full}$, effectively capturing deviations from the quantized values. 
To demonstrate how $\sigma_{xy}^{\rm free}$ can take non-quantized values, we plot the Berry curvature of the top ($\mu=+$) and bottom ($\mu=-$) quasienergy sub-band of the $n=0$ Floquet level in the Brillouin zone at two different values of $\tilde{A}=0.58$ (left panels) and $2.10$ (right panels) for each value of $\Omega/\tau$.
The shaded and unshaded areas show where states are occupied and unoccupied by electrons, respectively.
}
\label{FIG3}
\end{figure*}

Figure~\ref{FIG3} shows the comparison between $\sigma^{\rm full}_{xy}$, $\sigma^{\rm free}_{xy}$, and $\sigma^{\rm TKNN}_{xy}$ as a function of $\tilde{A}$ for three different values of $\Omega/\tau$.
As shown by the comparison, $\sigma_{xy}^{\rm free}$ closely matches $\sigma_{xy}^{\rm full}$, accurately capturing deviations from the quantized values.
This means that $\sigma_{xy}^{\rm free}$ can provide an intuitive understanding of how $\sigma_{xy}^{\rm full}$ can take non-quantized values.
The left and right panels of Fig.~\ref{FIG3} plots the Berry curvature of the top ($\mu=+$) and bottom ($\mu=-$) quasienergy sub-band of the $n=0$ Floquet level, ${\cal B}_{\mu=\pm,n=0,{\bf k}}$,  in the Brillouin zone, where occupied states, satisfying $f_{\rm FD}\left(\epsilon_{\mu=\pm,n=0,\mathbf{k}} - \langle \mathcal{N} \rangle_{\mu=\pm,n=0,\mathbf{k}} \Omega\right)=1$, are shown as shaded regions and unoccupied states, satisfying $f_{\rm FD}\left(\epsilon_{\mu=\pm,n=0,\mathbf{k}} - \langle \mathcal{N} \rangle_{\mu=\pm,n=0,\mathbf{k}} \Omega\right)=0$, as unshaded.
Note that $\sigma_{xy}^{\rm full}$ remains quantized when there is no population inversion, as shown in the left panels of Fig.~\ref{FIG3}~({\bf a}) at $\Omega/\tau=8$ and $\tilde{A}=0.58$.
By contrast, $\sigma_{xy}^{\rm full}$ deviates from the quantized values when population inversion occurs, as shown in the left panels of Fig.~\ref{FIG3}~({\bf b}) at $\Omega/\tau=4$ and $\tilde{A}=0.58$, and Fig.~\ref{FIG3}~({\bf c}) at $\Omega/\tau=2$ and $\tilde{A}=0.58$.
Meanwhile, in the right panels of Fig.~\ref{FIG3} at $\tilde{A}=2.10$, there is no population inversion, and thus $\sigma_{xy}^{\rm full}$ stays quantized across all three values of $\Omega/\tau$.

\begin{figure}[ht]
\includegraphics[width=\linewidth]{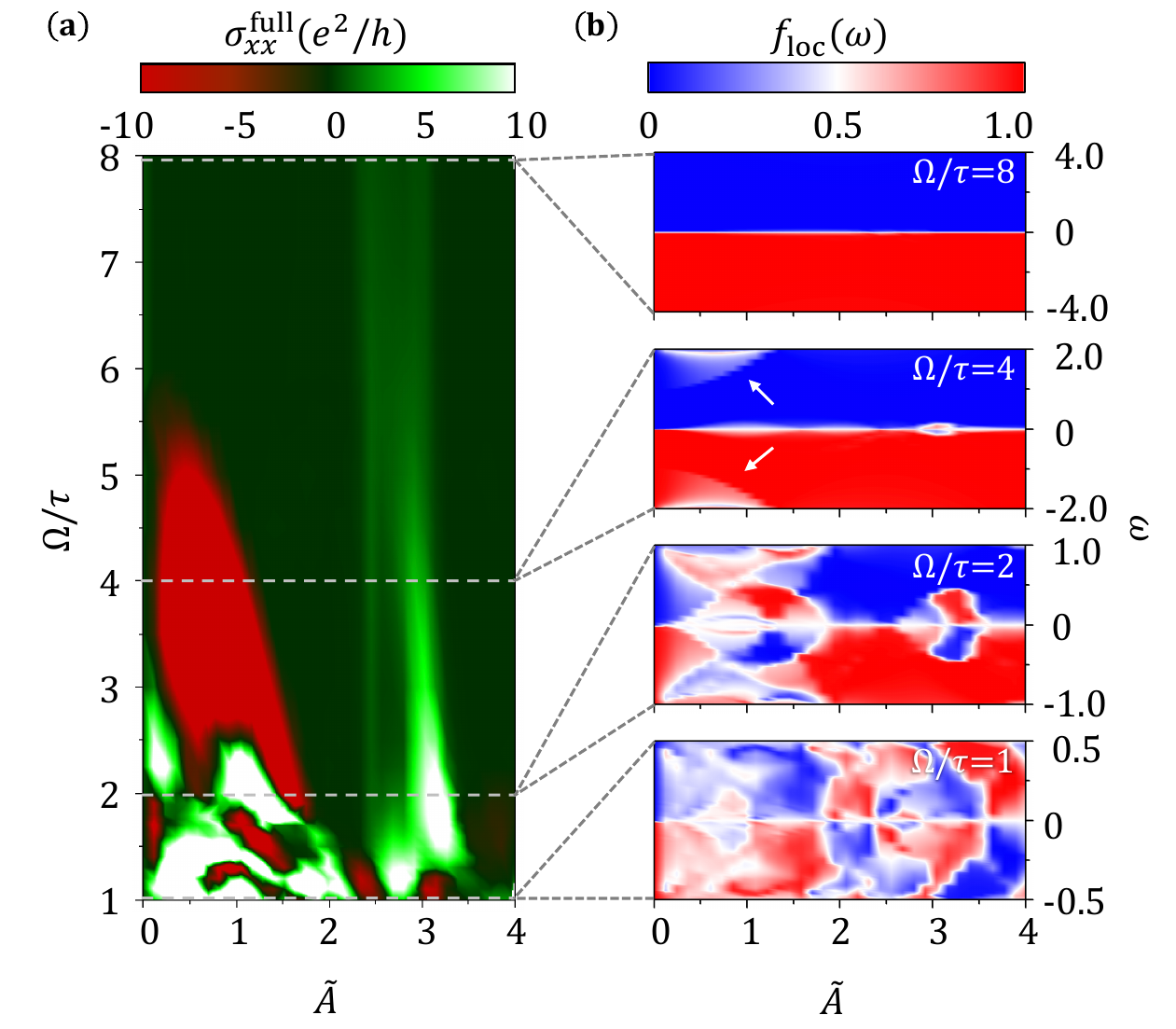}
\caption{\textbf{Negative longitudinal conductivity and population inversion.}
({\bf a}) The nonequilibrium longitudinal conductivity calculated from the fully interacting Green's functions, $\sigma^{\rm full}_{xx}$, is plotted in color as a function of $\Omega/\tau$ and $\tilde{A}$.
The dashed lines represent the constant frequency lines at $\Omega/\tau=8$, $4$, $2$, and $1$. 
({\bf b}) Local distribution function, $f_{\rm loc}(\omega)$, is plotted within the first Floquet Brillouin zone ($-\Omega/2<\omega\leq\Omega/2$) in color as a function of $\omega$ and $\tilde{A}$ at four different values of $\Omega/\tau$ corresponding to the dashed lines in ({\bf a}). 
The red and blue colors indicate the occupied and unoccupied states, respectively. 
}
\label{FIG4}
\end{figure}

Fourth, the orange curve resulting from the overshoot of $\sigma_{xy}^{\rm full}$ in Fig.~\ref{FIG2}~({\bf b}) aligns with the boundary of the regime where $\sigma_{xx}^{\rm full}$ is negative in Fig.~\ref{FIG2}~({\bf c}).
%Negative longitudinal conductivity triggers the NRC. 
To elucidate the relationship between negative longitudinal conductivity and population inversion, we present the local distribution function, $f_{\rm loc}(\omega)$, in Fig.~\ref{FIG4}.
At $\Omega/\tau=8$, $f_{\rm loc}(\omega)$ is identical to the Fermi-Dirac distribution function within the Floquet Brillouin zone regardless of the values of $\tilde{A}$.
At $\Omega/\tau=4$, population inversion begins to develop at the edges of the Floquet Brillouin zone at the values of $\tilde{A}$ indicated by the white arrows, exactly where the nonequilibrium longitudinal conductivity becomes negative. 
At $\Omega/\tau=2$ and $1$, $f_{\rm loc}(\omega)$ exhibits highly irregular behavior intertwined with population inversion, causing similarly irregular behavior in $\sigma_{xx}^{\rm full}$.

Fifth, to better understand the relationship between negative longitudinal conductivity and population inversion, we simplify the nonequilibrium Kubo formula for longitudinal conductivity in the fast-driving limit as follows:
%%%%%%%%%%%%%%%%%%%%%%%%%%%%%%%%%%%%%%%%%%%%%%%
\begin{align}
\sigma_{xx}^{\rm fast} =  2\pi^2 \int \frac{d^{2}\mathbf{k}}{(2\pi)^2} \int \frac{d\omega}{2\pi} 
\sum_n \mathrm{Tr}^\prime 
\left[\hat{J}^x_{\bf k} \rho_{\bf k}(\omega) \hat{J}^x_{\bf k} \rho_{\bf k}({\omega}) \right]_{nn}
\left(-\frac{\partial f_{{\rm loc},n}(\omega)}{\partial \omega}\right),
\label{eq:sigma_xx_fast}
\end{align}
%%%%%%%%%%%%%%%%%%%%%%%%%%%%%%%%%%%%%%%%%%%%%%%
where ${\rm Tr}^\prime$ indicates that the trace is performed over the sublattice indices, 
%$\hat{J}^x_{\bf k}$ is the Floquet current matrix defined in Eq.~\eqref{eq:J_hat_Floquet}, 
$\hat{J}^x_{\bf k}$ represents the Floquet matrix of the current operator, 
%expressed as a $2 \times 2$ matrix with respect to the sublattice index, 
$\rho_{\bf k}(\omega)$ is the Floquet matrix of the spectral function, and $f_{{\rm loc},n}(\omega)$ is the Floquet decomposition of the local distribution function, i.e., $f_{{\rm loc},n}(\omega)=f_{\rm loc}(\omega+n\Omega)$ for $-\Omega/2<\omega\leq\Omega/2$.
See Appendix~\ref{appen:sigma_xx_fast} for a detailed derivation of Eq.~\eqref{eq:sigma_xx_fast}.

\begin{figure*}[ht]
\includegraphics[width=\linewidth]{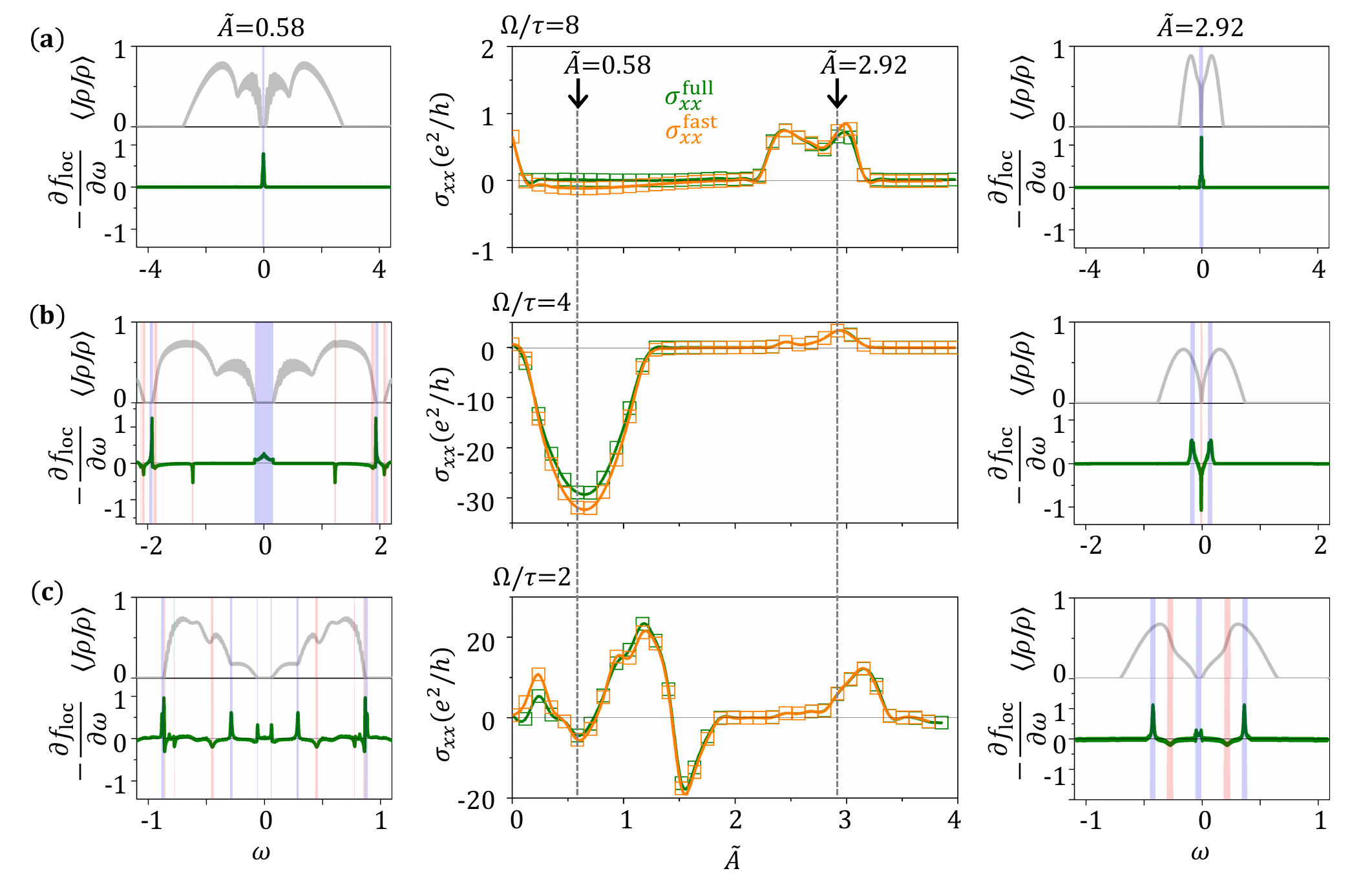}
\caption{
\textbf{Nonequilibrium longitudinal conductivity of irradiated graphene.}
The nonequilibrium longitudinal conductivity calculated from the fully interacting Green's functions, $\sigma_{xx}^{\rm full}$, is compared with that in the fast-driving limit, $\sigma_{xx}^{\rm fast}$. 
The comparison is conducted at three driving frequencies of ({\bf a}) $\Omega/\tau=8$, ({\bf b}) $\Omega/\tau=4$, and ({\bf c}) $\Omega/\tau=2$.
As shown in the comparison, $\sigma_{xx}^{\rm fast}$ closely matches $\sigma_{xx}^{\rm full}$ regardless of the values of $\tilde{A}$ across all three values of $\Omega/\tau$. 
Considering that $\sigma_{xx}^{\rm fast}$ is the frequency integral of the product of two factors, $\langle {J\rho J\rho} \rangle$ and $-\frac{\partial f_{\rm loc}}{\partial \omega}$, we plot $\langle {J\rho J\rho} \rangle$ and $-\frac{\partial f_{\rm loc}}{\partial \omega}$ in
the left and right panels for $\tilde{A}=0.58$ and $2.92$, respectively. 
We focus on the frequency range of the first Floquet Brillouin zone, $-\Omega/2 < \omega \leq \Omega/2$, because the main contribution to conductivity comes from this range.
Note that $\langle J\rho J\rho \rangle$  and $-\frac{\partial f_{\rm loc}}{\partial \omega}$ are plotted in properly normalized units because we are only interested in their relative strength as a function of $\omega$.
The red and blue shaded areas indicate that $-\frac{\partial f_{\rm loc}}{\partial \omega}<0$ and $-\frac{\partial f_{\rm loc}}{\partial \omega}>0$, respectively.
}
\label{FIG5}
\end{figure*}

Figure~\ref{FIG5} shows that $\sigma_{xx}^{\rm fast}$ closely matches $\sigma_{xx}^{\rm full}$ regardless of the values of $\tilde{A}$ across all three values of $\Omega/\tau=8$, $4$, and $2$,  indicating that $\sigma_{xx}^{\rm fast}$ captures the essence of $\sigma_{xx}^{\rm full}$ in a wide range of parameters.
Equation~\eqref{eq:sigma_xx_fast} shows that $\sigma_{xx}^{\rm fast}$ is the frequency integral of the product of two factors, $\langle {J\rho J\rho} \rangle = \int d^2 \mathbf{k} \mathrm{Tr}^{\prime} \left[ \hat{J}^{x}_{\mathbf{k}} \rho_\mathbf{k}(\omega) \hat{J}^{x}_{\mathbf{k}} \rho_\mathbf{k}(\omega)\right]_{nn}$ and $-\frac{\partial f_{\rm loc}}{\partial \omega}$.
Since $\langle {J\rho J\rho} \rangle$ is always positive, $\sigma_{xx}^{\rm fast}$ can become negative only when $-\frac{\partial f_{\rm loc}}{\partial \omega}<0$, i.e., when population inversion occurs. 
%The left and right panels of Fig.~\ref{FIG5} plot $\langle {J\rho J\rho} \rangle$ and $-\frac{\partial f_{\rm loc}}{\partial \omega}$ at $\tilde{A}=0.58$ and $2.92$, respectively. 
Figure~\ref{FIG5}~({\bf a}) shows that, at $\Omega/\tau=8$, there is no population inversion at any values of $\tilde{A}$, and therefore $\sigma_{xx}^{\rm fast}$ remains non-negative.
The left panels of Fig.~\ref{FIG5}~({\bf b}) and ({\bf c}) show that $\sigma_{xx}^{\rm fast}$ can become negative when population inversion occurs with the right combination of $\langle {J\rho J\rho} \rangle$.
By contrast, the right panels of Fig.~\ref{FIG5}~({\bf b}) and ({\bf c}) show that $\sigma_{xx}^{\rm fast}$ can return to positive values since the effect of population inversion is cancelled by contributions from even larger positive values of $-\frac{\partial f_{\rm loc}}{\partial \omega}$, resulting in net positive values for $\sigma_{xx}^{\rm fast}$.

As previously mentioned, negative longitudinal conductivity is predicted to trigger NRC, which induces a zero-resistance state with spontaneous formation of an inhomogeneous current distribution through the same mechanism as the quantum RIZR state~\cite{Andreev1,Auerbach1,Finkler1,Dmitriev1,Dmitriev2,Dorozhkin2,Wang2}.
So far, we have discussed when and how negative longitudinal conductivity can occur in irradiated graphene.
A natural question is how robust NRC can be against impurity scattering, which is expected to favor a more traditional Drude-like behavior.
We have confirmed that negative longitudinal conductivity and, consequently, NRC are robust against impurity scattering across a wide range of driving frequencies near half of the graphene band width, i.e., $\Omega \simeq {\cal W}/2$.
Refer to Appendix~\ref{appen:disorder_effect} for details.

\subsection*{Discussion}

In this work, we show that irradiated graphene can support two types of zero-resistance states when exposed to circularly polarized light: (i) Floquet Chern insulators and (ii) the NRC-induced RIZR state with spontaneous formation of an inhomogeneous current distribution.

Although both types of states exhibit zero resistance, they differ in the underlying mechanisms responsible for this phenomenon.  
Floquet Chern insulators exhibit zero resistance due to the topological protection of edge states in Floquet Chern bands. 
The NRC-induced RIZR state exhibits zero resistance due to the spontaneous formation of an inhomogeneous current distribution, caused by the S-shaped current-voltage characteristic curve.

Setting aside the underlying mechanisms, is there any experimental difference that can distinguish these two states?
One of the main differences is the current distribution.
The NRC-induced RIZR state has an inhomogeneous current distribution without any external voltage bias, which can, in principle, be directly measured in experiments. 
By contrast, Floquet Chern insulators are expected to have no current flowing through the bulk because they are indeed Chern insulators.
Another key difference is how the nonequilibrium anomalous Hall conductivity behaves as the electron density changes from half-filling.
Specifically, Floquet Chern insulators are expected to sustain quantized anomalous Hall conductivity, whereas the NRC-induced RIZR state exhibits unquantized values that vary with electron density.

We now assess the experimental feasibility by estimating the parameter ranges for Floquet Chern insulators and the NRC-induced RIZR state in irradiated graphene. 
If we assume that the graphene bandwidth ${\cal W}=6\tau$ is approximately 10 eV, Floquet Chern insulators would occur at similar or higher driving frequencies of light, corresponding to UV light.
Similarly, the NRC-induced RIZR state would occur at lower driving frequencies of light, corresponding to somewhere between visible and UV light.
Regarding the required electric-field strength of light, both Floquet Chern insulators and the NRC-induced RIZR state do not require too large electric fields.
For example, $\tilde{A}$ can be chosen to be any value close to unity, corresponding to electric-field strengths on the order of $\sim$ 0.1 V \AA$^{-1}$ for $\Omega \lesssim {\cal W}$, which are within experimentally accessible regimes~\cite{Merboldt1,Schiffrin1,Mayer1,Higuchi1,Uchida1,Li1}.

Finally, it would be interesting to explore the possibility of radiation-induced superconductivity. 
%one of the most important forms of zero resistance, by irradiating non-superconducting materials. 
One possibility is to flatten Floquet bands so that the relative interaction strengths surpass the band width, forming a Hubbard-like model~\cite{Castro1}.
In fact, various quantum materials can be created by manipulating the Floquet bands, a process dubbed Floquet engineering~\cite{Oka1,Rudner1,Shan1,Zhou1,Sandholzer1,Castro1,Kobayashi1,Zhou2,Takahashi1}.

% Produces the bibliography via BibTeX.
\
\\

\
\\
\textbf{Acknowledgements}

This work is supported in part by the KIAS Individual Grant, PG032303 (K.P.) and PG088602 (Y.K.).
Additionally, Y.K. is supported by a grant from the National Research Foundation of Korea (NRF), funded by the Ministry of Science and ICT of the Korean government (RS-2025-00553820).
The authors also thank the Center for Advanced Computation (CAC) at KIAS for providing the computing resources used in this work.

\
\\
\textbf{Author contributions}

\
\\
\textbf{Additional information}

%\textbf{Supplementary Information}

\textbf{Competing financial interests} 

The authors declare no competing financial interests.

%%% Figures %%%

\pagebreak

%%%%%%%%%%%%%%%%%%%%%%%%%%%%%%%%%%%%%%%%%%%%%%%%%%%%%%%%%%%%%%%%%%%%%
%%%%%%%%%%%%%%%%%%%%%%%%%%%%%%%%%%%%%%%%%%%%%%%%%%%%%%%%%%%%%%%%%%%%%

\pagebreak

%%%%%%%%%%%%%%%%%%%%%%%%%%%%%%%%%%%%%%%%%%%%%%%%%%%%%%%%%%%%%%%%%%%%%
%%%%%%%%%%%%%%%%%%%%%%%%%%%%%%%%%%%%%%%%%%%%%%%%%%%%%%%%%%%%%%%%%%%%%

\pagebreak

%%%%%%%%%%%%%%%%%%%%%%%%%%%%%%%%%%%%%%%%%%%%%%%%%%%%%%%%%%%%%%%%%%%%%
%%%%%%%%%%%%%%%%%%%%%%%%%%%%%%%%%%%%%%%%%%%%%%%%%%%%%%%%%%%%%%%%%%%%%

\pagebreak

%%%%%%%%%%%%%%%%%%%%%%%%%%%%%%%%%%%%%%%%%%%%%%%%%%%%%%%%%%%%%%%%%%%%%
%%%%%%%%%%%%%%%%%%%%%%%%%%%%%%%%%%%%%%%%%%%%%%%%%%%%%%%%%%%%%%%%%%%%%

\pagebreak

%%%%%%%%%%%%%%%%%%%%%%%%%%%%%%%%%%%%%%%%%%%%%%%%%%%%%%%%%%%%%%%%%%%%%
%%%%%%%%%%%%%%%%%%%%%%%%%%%%%%%%%%%%%%%%%%%%%%%%%%%%%%%%%%%%%%%%%%%%%

\pagebreak

\clearpage
%%%%%%%%%%%%%%%%%%%%%%%%%%%%%%%%%%%%%%% title
\appendix
\counterwithin{equation}{section}
\counterwithin{figure}{section}

\renewcommand{\thesection}{\Alph{section}} 
\renewcommand{\theequation}{\thesection.\arabic{equation}}
\renewcommand{\thefigure}{\thesection.\arabic{figure}}

\setcounter{page}{1}
\appendix

\begin{center}
{\LARGE\bfseries Supplementary Information: Floquet Chern Insulators and Radiation-Induced Zero Resistance in Irradiated Graphene}\\[2em]

{\large Youngjae Kim and Kwon Park$^{\ast}$}\\[1em]

{\itshape School of Physics, Korea Institute for Advanced Study, Seoul 02455, Korea}
\end{center}

\vspace{2em}
%%%%%%%%%%%%%%%%%%%%%%%%%%%%%%%%%%%%%%%%%%%%%
\clearpage

%%%%%%%%%%%%%%%%%%%%%%%%%%%%%%%%%%%%
\section{Nonequilibrium Kubo formula in the Floquet representation}
\label{appen:noneq_Kubo}
%%%%%%%%%%%%%%%%%%%%%%%%%%%%%%%%%%%%

We start by expressing the Hamiltonian for irradiated graphene in second quantization form:
%%%%%%%%%%%%%%%%%%%%%%%%%%%%%%%%%%%%%%%%%%%%%%%%%%%%%%%%%%%%%%%%%%%%%%%%%%
\begin{equation}
H_{\mathbf{k}}(t) = \Delta_\mathbf{k}(t) b^{\dagger}_{\mathbf{k}} a_{\mathbf{k}} +\Delta^*_\mathbf{k}(t) a^{\dagger}_{\mathbf{k}} b_{\mathbf{k}}  ,
\end{equation}
%%%%%%%%%%%%%%%%%%%%%%%%%%%%%%%%%%%%%%%%%%%%%%%%%%%%%%%%%%%%%%%%%%%%%%%%%%
where $\Delta_\mathbf{k}(t) = -\tau \left(e^{i(\mathbf{k}-e\mathbf{A}(t)/c) \cdot \mathbf{a}_{1}} + e^{i(\mathbf{k}-e\mathbf{A}(t)/c)  \cdot \mathbf{a}_{2}} + e^{i(\mathbf{k}-e\mathbf{A}(t)/c) \cdot \mathbf{a}_{3}} \right)$ with $\mathbf{a}_1 = l(0,1)$, $\mathbf{a}_2 = -\frac{l}{2}(\sqrt{3},1)$, and $\mathbf{a}_3 = \frac{l}{2}(\sqrt{3},-1)$ with $\tau$ being the hopping parameter, $l$ being the distance between nearest-neighbor carbon atoms, and $\mathbf{A}(t) = A(\cos{\Omega t}, \sin{\Omega t})$ being the vector potential.
Here, $a_{\mathbf{k}}$ and $a^\dagger_{\mathbf{k}}$ are the annihilation and creation operators, respectively, for sublattice A.
Similarly, $b_{\mathbf{k}}$ and $b^\dagger_{\mathbf{k}}$ are the annihilation and creation operators, respectively, for sublattice B.

The DC electric conductivities can be determined from the time-ordered current-current correlation functions:
%%%%%%%%%%%%%%%%%%%%%%%%%%%%%%%%%%%%%%%%%%%%%%%%%
\begin{align}
\Pi_{\alpha\beta}(t_{1},t_{2}) = -i \langle \hat{T} {J}^{\alpha}(t_{1}) {J}^{\beta}(t_{2}) \rangle ,
\label{eq:Pi}
\end{align}
%%%%%%%%%%%%%%%%%%%%%%%%%%%%%%%%%%%%%%%%%%%%%%%%%
where the current operator is given as follows:
%%%%%%%%%%%%%%%%%%%%%%%%%%%%%%%%%%%%%%%%%%%%%%%%
\begin{align}
J^{\alpha}(t) = - \int \frac{d^2{\mathbf{k}}}{(2\pi)^2} \frac{\partial}{\partial {k}_{\alpha}} H_{\mathbf{k}}(t)
= - \int \frac{d^2{\mathbf{k}}}{(2\pi)^2}
\left(
\frac{\partial \Delta_\mathbf{k}(t)}{\partial {k}_{\alpha}}  b^{\dagger}_{\mathbf{k}}(t) a_{\mathbf{k}}(t) 
+\frac{\partial \Delta^*_\mathbf{k}(t)}{\partial {k}_{\alpha}} a^{\dagger}_{\mathbf{k}}(t) b_{\mathbf{k}}(t) 
\right),
\end{align}
%%%%%%%%%%%%%%%%%%%%%%%%%%%%%%%%%%%%%%%%%%%%%%%%
with $\alpha$ and $\beta$ indicating the spatial directions, which can be either $x$ or $y$.
Importantly, $a_{\mathbf{k}}(t)$ and $a^\dagger_{\mathbf{k}}(t)$ are the annihilation and creation operators in the Heisenberg picture, respectively, for sublattice A.
Specifically, $a_{\mathbf{k}}(t) = \hat{U}^{\dagger}(t,0)a_{\mathbf{k}}\hat{U}(t,0)$ and $a_{\mathbf{k}}^{\dagger}(t) = \hat{U}^{\dagger}(t,0)a_{\mathbf{k}}^{\dagger}\hat{U}(t,0)$, where $\hat{U}(t,0) = \hat{T} e^{-i \int_0^{t} dt' H_{\mathbf{k}}(t')}$ is the time evolution operator.  
Similar expressions hold for $b_{\mathbf{k}}(t)$ and $b^\dagger_{\mathbf{k}}(t)$.

In principle, the current-current correlation functions in Eq.~\eqref{eq:Pi} must be calculated for the fully interacting ground state in the presence of impurity scattering.
However, it is not feasible to account for both self-energy and vertex corrections caused by impurity scattering due to technical difficulties.
Therefore, in this work, we only consider self-energy corrections within the framework of the self-consistent Born approximation (SCBA), assuming that impurity scattering is sufficiently weak.
This means that we first calculate the current-current correlation functions for the noninteracting ground state and then later substitute the noninteracting Green's functions with the fully interacting Green's functions.

Using Wick's theorem, the noninteracting time-ordered current-current correlation functions can be rewritten as follows:
%%%%%%%%%%%%%%%%%%%%%
\begin{align}
\Pi^{(0)}_{\alpha\beta}(t_{1},t_{2}) 
= -i \int \frac{d^2 {\bf k}}{(2\pi)^2}
\textrm{Tr}^\prime
\left( 
\hat{J}^{\alpha}_\mathbf{k}(t_{1})
\hat{G}_{0,{\mathbf{k}}}(t_{1},t_{2})
\hat{J}^{\beta}_\mathbf{k}(t_{2})  
\hat{G}_{0,{\mathbf{k}}}(t_{2},t_{1})
\right),
\end{align}
%%%%%%%%%%%%%%%%%%%%%
where ${\rm Tr}^\prime$ indicates that the trace is performed over the sublattice indices. 
Here, $\hat{J}^\alpha_{\bf k}(t)$ is the current matrix defined as
%%%%%%%%%%%%%%%%%%%%%%%%%%%%%%
\begin{align}
\hat{J}^\alpha_{\bf k}(t)
=
\begin{pmatrix}
0 & -\frac{\partial \Delta_{\mathbf{k}}(t)}{\partial k^\alpha} \\
-\frac{\partial \Delta_{\mathbf{k}}^*}{\partial k^\alpha} & 0
\end{pmatrix} ,
\label{eq:J_hat}
\end{align}
%%%%%%%%%%%%%%%%%%%%%%%%%%%%%%
and $\hat{G}_{0,{\bf k}}(t,t^\prime)$ is the noninteracting Green's function matrix in the absence of impurity scattering, defined as 
%%%%%%%%%%%%%%%%%%%%%%%%%%%%%%
\begin{align}
\hat{G}_{0,{\bf k}}(t,t^\prime)
=
\begin{pmatrix}
G_{0,{\bf k},aa}(t,t^\prime) & G_{0,{\bf k},ab}(t,t^\prime) \\
G_{0,{\bf k},ba}(t,t^\prime) & G_{0,{\bf k},bb}(t,t^\prime)
\end{pmatrix} ,
\label{eq:G0_hat}
\end{align}
%%%%%%%%%%%%%%%%%%%%%%%%%%%%%%
where $G_{0,\mathbf{k},aa}(t,t') = -i\langle \hat{T} a_{\mathbf{k}}(t) a_{\mathbf{k}}^{\dagger}(t') \rangle$, $G_{0,\mathbf{k},bb}(t,t') = -i\langle \hat{T} b_{\mathbf{k}}(t) b_{\mathbf{k}}^{\dagger}(t') \rangle$, $G_{0,\mathbf{k},ab}(t,t') = -i\langle \hat{T} a_{\mathbf{k}}(t) b_{\mathbf{k}}^{\dagger}(t') \rangle$, and $G_{0,\mathbf{k},ba}(t,t') = -i\langle \hat{T} b_{\mathbf{k}}(t) a_{\mathbf{k}}^{\dagger}(t') \rangle$.

Next, we convert the time-ordered current-current correlation functions into their retarded counterparts using Langreth's theorem:
%%%%%%%%%%%%%%%%%%%%%%%%%%%%%%%%%%%%%%%%%%%%%%
\begin{align}
G(t_1,t_2)G(t_2,t_1) \rightarrow G^{r}(t_1,t_2)G^{<}(t_2,t_1) + G^{<}(t_1,t_2)G^{a}(t_2,t_1),
\end{align}
%%%%%%%%%%%%%%%%%%%%%%%%%%%%%%%%%%%%%%%%%%%%%%
which enables us to express the noninteracting retarded current-current correlation functions in terms of the noninteracting retarded and lesser Green's functions as follows:
%%%%%%%%%%%%%%%%%%%%%%%%%%%%%%%%%%%%%%%%%%%%%%%%%%%%%%%%%%%%%%%%%%%%%%%%%%%%%
\begin{align}
\Pi^{(0) r}_{\alpha\beta}(t_{1},t_{2}) = 
-i \int \frac{d^2 {\bf k}}{(2\pi)^2}
\textrm{Tr}^\prime \Big( 
&\hat{J}^{\alpha}_\mathbf{k}(t_{1}) \hat{G}^{r}_{0,\mathbf{k}}(t_{1},t_{2}) \hat{J}^{\beta}_\mathbf{k}(t_{2}) \hat{G}^{<}_{0,\mathbf{k}}(t_{2},t_{1})
\nonumber \\
&+ \hat{J}^{\alpha}_\mathbf{k}(t_{1}) \hat{G}^{<}_{0,\mathbf{k}}(t_{1},t_{2}) \hat{J}^{\beta}_\mathbf{k}(t_{2}) \hat{G}^{a}_{0,\mathbf{k}}(t_{2},t_{1}) 
\Big).
\end{align}
%%%%%%%%%%%%%%%%%%%%%%%%%%%%%%%%%%%%%%%%%%%%%%%%%%%%%%%%%%%%%%%%%%%%%%%%%%%%%

Now, it is convenient to transform the retarded current-current correlation functions into the frequency domain, which can be accomplished through the Floquet Fourier transform:
%%%%%%%%%%%%%%%%%%%%%%%%%%%%%%%%%%%%%%%%%%%%%%%%%%%%%%%%%%%%%%%%%%%%
\begin{align}
[\xi(\omega)]_{nm} = \int dt \int dt^\prime e^{i(\omega+n\mathit\Omega)t} \xi(t,t^\prime) e^{-i(\omega+m\mathit\Omega)t^\prime},
\end{align}
%%%%%%%%%%%%%%%%%%%%%%%%%%%%%%%%%%%%%%%%%%%%%%%%%%%%%%%%%%%%%%%%%%%%
where $\xi(t,t^\prime)$ denotes any general correlation function.
The Floquet Fourier transform is complemented by the inverse Floquet Fourier transform:
%%%%%%%%%%%%%%%%%%%%%%%%%%%%%%%%%%%%%%%%%%%%%%%%%%%%%%%%%%%%%%%%%%%%
\begin{align}
\xi(t,t^\prime)= \int \frac{d\omega}{2\pi} \sum_{n,m} e^{-i(\omega+n\mathit\Omega)t} [\xi(\omega)]_{nm} e^{i(\omega+m\mathit\Omega)t^\prime}.
\end{align}
%%%%%%%%%%%%%%%%%%%%%%%%%%%%%%%%%%%%%%%%%%%%%%%%%%%%%%%%%%%%%%%%%%%%
We can apply the Floquet Fourier transform to the retarded and lesser Green's functions as follows:
%%%%%%%%%%%%%%%%%%%%%%%%%%%%%%%%%%%%%%%%%%%%%%%%%%%%%%%%%%%%%%%%%%%%
\begin{align}
[\hat{G}^{r,<}({\bf k},\omega)]_{nm} = \int dt \int dt^\prime e^{i(\omega+n\mathit\Omega)t} 
\hat{G}^{r,<}_{\bf k}(t,t^\prime) e^{-i(\omega+m\mathit\Omega)t^\prime},
\end{align}
%%%%%%%%%%%%%%%%%%%%%%%%%%%%%%%%%%%%%%%%%%%%%%%%%%%%%%%%%%%%%%%%%%%%
which is complemented by the inverse Floquet Fourier transform:
%%%%%%%%%%%%%%%%%%%%%%%%%%%%%%%%%%%%%%%%%%%%%%%%%%%%%%%%%%%%%%%%%%%%
\begin{align}
\hat{G}^{r,<}_{\bf k}(t,t^\prime)= \int \frac{d\omega}{2\pi} \sum_{n,m} e^{-i(\omega+n\mathit\Omega)t} 
[\hat{G}^{r,<}({\bf k},\omega)]_{nm} e^{i(\omega+m\mathit\Omega)t^\prime}.
\end{align}
%%%%%%%%%%%%%%%%%%%%%%%%%%%%%%%%%%%%%%%%%%%%%%%%%%%%%%%%%%%%%%%%%%%%
Meanwhile, the current operator can be transformed using the usual Fourier transform as follows:
%%%%%%%%%%%%%%%%%%%%%%%%%%%%%%%%%%%%%%%%%%%%%%%%
\begin{align}
\tilde{J}^{\alpha}_{{\mathbf{k}},l} =  \int dt \hat{J}^\alpha_{\bf k}(t) e^{il\Omega t},
\end{align}
%%%%%%%%%%%%%%%%%%%%%%%%%%%%%%%%%%%%%%%%%%%%%%%%
which is also complemented by the usual inverse Fourier transform:
%%%%%%%%%%%%%%%%%%%%%%%%%%%%%%%%%%%%%%%%%%%%%%%%
\begin{align}
\hat{J}^{\alpha}_{\mathbf{k}}(t) = \sum_l \tilde{J}^\alpha_{{\bf k},l} e^{-il\Omega t}.
\end{align}
%%%%%%%%%%%%%%%%%%%%%%%%%%%%%%%%%%%%%%%%%%%%%%%%
Bringing them all together, we can transform the noninteracting retarded current-current correlation functions into the frequency domain as follows:
%%%%%%%%%%%%%%%%%%%%%%%%%%%%%%%%%%%%%%%%%%%%%%%
\begin{align}
[\Pi^{(0) r}_{\alpha\beta}(\omega)]_{nm}=
-i \int \frac{d^2{\bf k}}{(2\pi)^2} \int \frac{d\bar{\omega}}{2\pi} 
\sum_{l_1,l_2,j_1,j_2}
\textrm{Tr}^\prime
\Big(
&\tilde{J}^{\alpha}_{\mathbf{k},l_1}
\left[\hat{G}^{r}_{0}({\bf k},\bar{\omega}+\omega)\right]_{\substack{j_2-l_1+n,\\j_1+l_2+m}} 
\tilde{J}^{\beta}_{\mathbf{k},l_2}
\left[\hat{G}^{<}_{0}({\bf k},\bar{\omega})\right]_{j_1j_2} 
\nonumber \\ 
+&\tilde{J}^{\alpha}_{\mathbf{k},l_1} 
\left[\hat{G}^{<}_{0}({\bf k},\bar{\omega})\right]_{j_1j_2} 
\tilde{J}^{\beta}_{\mathbf{k},l_2}
\left[\hat{G}^{a}_{0}({\bf k},\bar{\omega}-\omega)\right]_{\substack{j_2-l_2-m,\\j_1+l_1-n}} 
\Big),
\end{align}
%%%%%%%%%%%%%%%%%%%%%%%%%%%%%%%%%%%%%%%%%%%%%%%%

Since our primary interest in this work is the DC conductivities in the zero-frequency limit, we now focus on the $(n,m) = (0,0)$ component of the current-current correlation functions:
%%%%%%%%%%%%%%%%%%%%%%%%%%%%%%%%%%%%%%%%%%%%%%%
\begin{align}
[\Pi^{(0) r}_{\alpha\beta}(\omega)]_{00}=
-i \int \frac{d^2{\bf k}}{(2\pi)^2} \int \frac{d\bar{\omega}}{2\pi} 
\sum_{l_1,l_2,j_1,j_2}
\textrm{Tr}^\prime
\Big(
&\tilde{J}^{\alpha}_{\mathbf{k},l_1}
\left[\hat{G}^{r}_{0}({\bf k},\bar{\omega}+\omega)\right]_{\substack{j_2-l_1,\\j_1+l_2}} 
\tilde{J}^{\beta}_{\mathbf{k},l_2}
\left[\hat{G}^{<}_{0}({\bf k},\bar{\omega})\right]_{j_1j_2} 
\nonumber \\ 
+&\tilde{J}^{\alpha}_{\mathbf{k},l_1} 
\left[\hat{G}^{<}_{0}({\bf k},\bar{\omega})\right]_{j_1j_2} 
\tilde{J}^{\beta}_{\mathbf{k},l_2}
\left[\hat{G}^{a}_{0}({\bf k},\bar{\omega}-\omega)\right]_{\substack{j_2-l_2,\\j_1+l_1}} 
\Big),
\end{align}
%%%%%%%%%%%%%%%%%%%%%%%%%%%%%%%%%%%%%%%%%%%%%%%%
which can be further simplified using the Toeplitz property of the Floquet matrix as follows:
%%%%%%%%%%%%%%%%%%%%%%%%%%%%%%%%%%%%%%%%%%%%%%%
\begin{align}
[\Pi^{(0) r}_{\alpha\beta}(\omega)]_{00}=
-i \int \frac{d^2{\bf k}}{(2\pi)^2} \int \frac{d\bar{\omega}}{2\pi} 
\sum_{l_1,l_2,j_1,j_2}
\textrm{Tr}^\prime
\Big(
&[\hat{J}^{\alpha}_{\mathbf{k}}]_{l_1 j_1}
\left[\hat{G}^{r}_{0}({\bf k},\bar{\omega}+\omega)\right]_{j_1 l_2}
[\hat{J}^{\beta}_{\mathbf{k}}]_{l_2 j_2}
\left[\hat{G}^{<}_{0}({\bf k},\bar{\omega})\right]_{j_2 l_1} 
\nonumber \\ 
&+[\hat{J}^{\alpha}_{\mathbf{k}}]_{l_1 j_1} 
\left[\hat{G}^{<}_{0}({\bf k},\bar{\omega})\right]_{j_1 l_2} 
[\hat{J}^{\beta}_{\mathbf{k}}]_{l_2 j_2}
\left[\hat{G}^{a}_{0}({\bf k},\bar{\omega}-\omega)\right]_{j_2 l_1} 
\Big),
\label{eq:Pi_Toeplitz}
\end{align}
%%%%%%%%%%%%%%%%%%%%%%%%%%%%%%%%%%%%%%%%%%%%%%%%
where the Floquet current matrix is defined as
%%%%%%%%%%%%%%%%%%%%%%%%%%%%%%%%%%%%%%%%%%%%%%%
\begin{align}
[\hat{J}^{\alpha}_{\mathbf{k}}]_{nm}=\tilde{J}^\alpha_{{\bf k},n-m}.
\label{eq:J_hat_Floquet}
\end{align}
%%%%%%%%%%%%%%%%%%%%%%%%%%%%%%%%%%%%%%%%%%%%%%%%

It is noteworthy that the subscripts in Eq.~\eqref{eq:Pi_Toeplitz} align naturally, making the entire expression in the integrand correspond to the trace operation over the Floquet matrices. 
Consequently, Equation~\eqref{eq:Pi_Toeplitz} can be simplified as follows:
%%%%%%%%%%%%%%%%%%%%%%%%%%%%%%%%%%%%%%%%%%%%%%%
\begin{align}
[\Pi^{(0) r}_{\alpha\beta}(\omega)]_{00}=
-i \int \frac{d^2{\bf k}}{(2\pi)^2} \int \frac{d\bar{\omega}}{2\pi} 
\textrm{Tr}
\Big(
{J}^{\alpha}_{\mathbf{k}}
G^{r}_{0}({\bf k},\bar{\omega}+\omega)
{J}^{\beta}_{\mathbf{k}}
G^{<}_{0}({\bf k},\bar{\omega})
+{J}^{\alpha}_{\mathbf{k}}
G^{<}_{0}({\bf k},\bar{\omega})
{J}^{\beta}_{\mathbf{k}}
G^{a}_{0}({\bf k},\bar{\omega}-\omega)
\Big),
\label{eq:Pi_trace}
\end{align}
%%%%%%%%%%%%%%%%%%%%%%%%%%%%%%%%%%%%%%%%%%%%%%%%
where $J^\alpha_{\bf k}$ and $G^{r,<}({\bf k},\omega)$ represent the current operator matrix and the Green's function matrices, respectively, expressed with combined indices for both sublattices and Floquet components.
Here, $\textrm{Tr}$ indicates that the trace is performed over both sublattice and Floquet indices. 
As mentioned previously, in this work, we calculate the fully interacting current-current correlation functions using the SCBA framework, which corresponds to substituting the noninteracting Green's functions with the fully interacting Green's functions in Eq.~\eqref{eq:Pi_trace}:
%Therefore, we reach the final expression for the fully interacting retarded current-current correlation functions as follows:
%%%%%%%%%%%%%%%%%%%%%%%%%%%%%%%%%%%%%%%%%%%%%%%
\begin{align}
[\Pi^{r}_{\alpha\beta}(\omega)]_{00}=
-i \int \frac{d^2{\bf k}}{(2\pi)^2} \int \frac{d\bar{\omega}}{2\pi} 
\textrm{Tr}
\Big(
{J}^{\alpha}_{\mathbf{k}}
G^{r}({\bf k},\bar{\omega}+\omega)
{J}^{\beta}_{\mathbf{k}}
G^{<}({\bf k},\bar{\omega})
+{J}^{\alpha}_{\mathbf{k}}
G^{<}({\bf k},\bar{\omega})
{J}^{\beta}_{\mathbf{k}}
G^{a}({\bf k},\bar{\omega}-\omega)
\Big),
\label{eq:Pi_final}
\end{align}
%%%%%%%%%%%%%%%%%%%%%%%%%%%%%%%%%%%%%%%%%%%%%%%%

The DC electric conductivities are defined in the zero-frequency limit as follows:
%%%%%%%%%%%%%%%%%%%%%%%%%%%%%%%%%%%%%%%%%%%%%%%%%%%%
\begin{align}
\sigma_{\alpha\beta} = -\lim_{\omega\to 0} \mathrm{Im} \, \mathrm{Tr} \left( \frac{\Pi^{r}_{\alpha\beta}(\omega)}{\omega} \right).
\end{align}
%%%%%%%%%%%%%%%%%%%%%%%%%%%%%%%%%%%%%%%%%%%%%%%%%%%%
Then, using the identity,
%%%%%%%%%%%%%%%%%%%%%%%%%%%%%%%%%%%%%%%%%%%%%%%%%%%%%%%%%
\begin{align}
\mathrm{Im} \, \mathrm{Tr} \left( \frac{\Pi^{r}_{\alpha\beta}(\omega)}{\omega} \right) = 
i \mathrm{Tr} \left( \frac{{\Pi^{r}_{\alpha\beta}(\omega)}^{\dagger} - \Pi^{r}_{\alpha\beta}(\omega)}{2\omega} \right),
\end{align}
%%%%%%%%%%%%%%%%%%%%%%%%%%%%%%%%%%%%%%%%%%%%%%%%%%%%%%%%%
we can obtain the DC conductivities in terms of the retarded and lesser Green's functions as follows:
%%%%%%%%%%%%%%%%%%%%%%%%%%%%%%%%%%%%%%%%%%%%%%%%
\begin{align}
\sigma_{\alpha\beta} 
= \lim_{\omega \rightarrow 0}
\int \frac{d^2{\bf k}}{(2\pi)^2} \int \frac{d\bar{\omega}}{2\pi} \mathrm{Tr} 
\Bigg( 
&{J}^{\alpha}_{\mathbf{k}} 
\frac{G^{r}({\bf k},\bar{\omega}+\omega) - G^{r}({\bf k},\bar{\omega}-\omega)}{2\omega}
{J}^{\beta}_{\mathbf{k}}
G^{<}({\bf k},\bar{\omega}) 
\nonumber \\
&-{J}^{\alpha}_{\mathbf{k}}
G^{<}({\bf k},\bar{\omega})
{J}^{\beta}_{\mathbf{k}}
\frac{G^{a}({\bf k},\bar{\omega}+\omega) - G^{a}({\bf k},\bar{\omega}-\omega)}{2\omega} 
\Bigg),
\end{align}
%%%%%%%%%%%%%%%%%%%%%%%%%%%%%%%%%%%%%%%%%%%%%%%%
where we have used the relations that $[G^{<}]^{\dagger}=-G^<$ and $[G^{r}]^{\dagger}=G^a$.

Recognizing that the zero-frequency limit amounts to the first derivative of the Green’s functions with respect to $\omega$, we arrive at the final expression for the DC electric conductivities:
%%%%%%%%%%%%%%%%%%%%%%%%%%%%%%%%%%%%%%%%%%%%%%%%%%%%%%%%%%%%%%%%%%%%%%%%%%%%
\begin{equation}
\sigma_{\alpha \beta} =  \int \frac{d^{2}\mathbf{k}}{(2\pi)^2} \int \frac{d\omega}{2\pi} 
\mathrm{Tr} 
\left( 
J^{\alpha}_{\mathbf{k}} \frac{\partial G^{r}({\mathbf{k}},\omega)}{\partial \omega} J^{\beta}_{\mathbf{k}} G^{<}(\mathbf{k},\omega) 
-J^{\alpha}_{\mathbf{k}} G^{<}(\mathbf{k},\omega) {J}^{\beta}_{\mathbf{k}}\frac{\partial G^{a}(\mathbf{k},\omega)}{\partial \omega}
\right),
\label{eq:sigma_final}
\end{equation}
%%%%%%%%%%%%%%%%%%%%%%%%%%%%%%%%%%%%%%%%%%%%%%%%%%%%%%%%%%%%%%%%%%%%%%%%%%%%
which is exactly the same expression as the one obtained previously~\cite{Mikami_Thesis}.

%%%%%%%%%%%%%%%%%%%%%%%%%%%%%%%%%%%%%%%%%%%%%%%%%%%%
\section{Nonequilibrium Kubo formula for anomalous Hall conductivity in the impurity-free limit}
\label{appen:sigma_xy_free}
%%%%%%%%%%%%%%%%%%%%%%%%%%%%%%%%%%%%%%%%%%%%%%%%%%%%

In the impurity-free limit, the nonequilibrium Kubo formula for anomalous Hall conductivity can be simplified as follows:
%%%%%%%%%%%%%%%%%%%%%%%%%%%%%%%%%%%%%%%%%%%%%%%%%%%%%%%%%%%%
\begin{equation}
\sigma_{xy}^{\rm free} =  \int \frac{d^{2}\mathbf{k}}{(2\pi)^2} \int \frac{d\omega}{2\pi} 
\mathrm{Tr} 
\left( 
J^x_{\mathbf{k}} \frac{\partial G^{r}_0({\mathbf{k}},\omega)}{\partial \omega} J^y_{\mathbf{k}} G^{<}_0(\mathbf{k},\omega) 
-J^x_{\mathbf{k}} G^{<}_0(\mathbf{k},\omega) {J}^y_{\mathbf{k}}\frac{\partial G^{a}_0(\mathbf{k},\omega)}{\partial \omega}
\right).
\label{eq:sigma_xy_free_0}
\end{equation}
%%%%%%%%%%%%%%%%%%%%%%%%%%%%%%%%%%%%%%%%%%%%%%%%%%%%%%%%%%%%

To start, using the unitary operator, $\mathcal{U}_{\mathbf{k}}$, which diagonalizes the Floquet Hamiltonian for irradiated graphene in the impurity-free limit, we can transform the operators within the integrand as follows:
%%%%%%%%%%%%%%%%%%%%%%%%%%%%%%%%%%%%%%%%%%%%%%%%%%%%%%%%%%%%
\begin{align}
\sigma_{xy}^{\rm free} = \int \frac{d^{2}\mathbf{k}}{(2\pi)^2} \int  \frac{d\omega}{2\pi}  
\mathrm{Tr} \Bigg(
&\left(\mathcal{U}_{\mathbf{k}}^{\dagger} {J}^{x}_{\mathbf{k}} \mathcal{U}_{\mathbf{k}}\right)
\left(\mathcal{U}_{\mathbf{k}}^{\dagger} \frac{\partial G^{r}_0({\mathbf{k}},\omega)}{\partial \omega} \mathcal{U}_{\mathbf{k}}\right)
\left(\mathcal{U}_{\mathbf{k}}^{\dagger} {J}^{y}_{\mathbf{k}} \mathcal{U}_{\mathbf{k}}\right)
\left(\mathcal{U}_{\mathbf{k}}^{\dagger} G^{<}_0({\mathbf{k}},\omega) \mathcal{U}_{\mathbf{k}}\right)
\nonumber \\
&-\left(\mathcal{U}_{\mathbf{k}}^{\dagger} {J}^{x}_{\mathbf{k}} \mathcal{U}_{\mathbf{k}}\right)
\left(\mathcal{U}_{\mathbf{k}}^{\dagger} G^{<}_0({\mathbf{k}},\omega) \mathcal{U}_{\mathbf{k}}\right)
\left(\mathcal{U}_{\mathbf{k}}^{\dagger} {J}^{y}_{\mathbf{k}} \mathcal{U}_{\mathbf{k}}\right)
\left(\mathcal{U}_{\mathbf{k}}^{\dagger} \frac{\partial G^{a}_0({\mathbf{k}},\omega)}{\partial \omega} \mathcal{U}_{\mathbf{k}}\right) 
\Bigg),
\label{eq:sigma_xy_unitary}
\end{align}
%%%%%%%%%%%%%%%%%%%%%%%%%%%%%%%%%%%%%%%%%%%%%%%%%%%%%%%%%%%%
where we have utilized the cyclic property of the ${\rm Tr}$ operator.

At this stage, it is crucial to remember that the unitary-transformed noninteracting Green's functions are simply the Green's functions of the Floquet eigenstate:
%%%%%%%%%%%%%%%%%%%%%%%%%%%%%%%%%%%%%%%%%%%%%%%%%%%%%%%%%%%%%
\begin{align}
{\cal G}_{\rm F}^{r,<}(\mathbf{k},\omega) = \mathcal{U}_{\mathbf{k}}^{\dagger} G^{r,<}_0({\mathbf{k}},\omega) \mathcal{U}_{\mathbf{k}},
\end{align}
%%%%%%%%%%%%%%%%%%%%%%%%%%%%%%%%%%%%%%%%%%%%%%%%%%%%%%%%%%%%%
%%%%%%%%%%%%%%%%%%%%%%%%%%%%%%%%%%%%%%%%%%%%%%%%%%%%%%%%%%%%%%
%\begin{align}
%G^{r,<}_{0}(\mathbf{k},\omega) = \mathcal{U}_{\mathbf{k}} g^{r,<}(\mathbf{k},\omega) \mathcal{U}^{\dagger}_{\mathbf{k}} ,
%\label{eq:G_0}
%\end{align}
%%%%%%%%%%%%%%%%%%%%%%%%%%%%%%%%%%%%%%%%%%%%%%%%%%%%%%%%%%%%%%
whose specific forms are given as follows:
%%%%%%%%%%%%%%%%%%%%%%%%%%%%%%%%%%%%%%%%%%%%%%%%%%%%%%%%%%%%%%%%%%%%%%%%
\begin{align}
[{\cal G}^{r}_{\rm F}(\mathbf{k},\omega)]_{\mu n,\nu m} &= \frac{1}{\omega - \epsilon_{\mu,n,\mathbf{k}} + i\delta}\delta_{\mu\nu}\delta_{nm}, \\
[{\cal G}^{<}_{\rm F}(\mathbf{k},(\omega)]_{\mu n,\nu m} &= 2\pi i\delta(\omega - \epsilon_{\mu,n,\mathbf{k}}) f_{\rm FD}(\omega - \langle \mathcal{N} \rangle_{\mu,n,\mathbf{k}} \Omega)\delta_{\mu\nu}\delta_{nm},
\end{align}
%%%%%%%%%%%%%%%%%%%%%%%%%%%%%%%%%%%%%%%%%%%%%%%%%%%%%%%%%%%%%%%%%%%%%%%%
where  $\epsilon_{\mu,n,\mathbf{k}}=\epsilon_{\mu,\mathbf{k}}+n\Omega$ and $\langle \mathcal{N} \rangle_{{\mu,n,\mathbf{k}}}=\langle \phi_{\mu,n,\mathbf{k}} | \mathcal{N} | \phi_{\mu,n,\mathbf{k}} \rangle$.
Note that $[{\mathcal{N}}]_{\mu\nu,nm}=n\delta_{\mu\nu}\delta_{nm}$ is the Floquet index operator, and $|\phi_{\mu,n,\mathbf{k}}\rangle$ is the Floquet eigenstate with quasienergy eigenvalue $\epsilon_{\mu,n,\mathbf{k}}$.
In a similar way, we can define the unitary-transformed current operator as the current operator of the Floquet eigenstate:
%%%%%%%%%%%%%%%%%%%%%%%%%%%%%%%%%%%%%%%%%%%%%%%%%%%%%%%%%
\begin{align}
\mathcal{J}^{x,y}_{\mathbf{k}} = \mathcal{U}_{\mathbf{k}}^{\dagger} {J}^{x,y}_{\mathbf{k}} \mathcal{U}_{\mathbf{k}},
\end{align}
%%%%%%%%%%%%%%%%%%%%%%%%%%%%%%%%%%%%%%%%%%%%%%%%%%%%%%%%%
which measures the current for each quasienergy band of the Floquet eigenstate.

Then, using ${\cal G}_{\rm F}^{r,<}(\mathbf{k},\omega)$ and $\mathcal{J}^{x,y}_{\mathbf{k}}$, we can rewrite Eq.~\eqref{eq:sigma_xy_unitary} as follows:
%%%%%%%%%%%%%%%%%%%%%%%%%%%%%%%%%%%%%%%%%%%%%%%%%%%%%%%%%%%%
\begin{align}
\sigma_{xy}^{\rm free} = \int \frac{d^{2}\mathbf{k}}{(2\pi)^2} \int  \frac{d\omega}{2\pi}  
\mathrm{Tr} \left(
\mathcal{J}^{x}_{\mathbf{k}}
\frac{\partial {\cal G}_{\rm F}^{r}({\mathbf{k}},\omega)}{\partial \omega}
\mathcal{J}^{y}_{\mathbf{k}}
{\cal G}_{\rm F}^{<}({\mathbf{k}},\omega)
-\mathcal{J}^{x}_{\mathbf{k}}
{\cal G}_{\rm F}^{<}({\mathbf{k}},\omega)
\mathcal{J}^{y}_{\mathbf{k}}
\frac{\partial {\cal G}_{\rm F}^{a}({\mathbf{k}},\omega)}{\partial \omega}
\right),
\end{align}
%%%%%%%%%%%%%%%%%%%%%%%%%%%%%%%%%%%%%%%%%%%%%%%%%%%%%%%%%%%%
which can be further simplified using the specific forms of ${\cal G}_{\rm F}^{<}(\mathbf{k},\omega)$ and $\partial {\cal G}_{\rm F}^{r,a}({\bf k},\omega)/\partial \omega$ as follows:
%%%%%%%%%%%%%%%%%%%%%%%%%%%%%%%%%%%%%%%%%%%%%%%%%%%%%%%%%%%%%%%%%%%%%%%%%%
\begin{align}
\sigma_{xy}^{\rm free} = 
-i \int \frac{d^{2}\mathbf{k}}{(2\pi)^2} 
\sum^\prime_{(\mu, n),(\nu, m)}  \frac{[\mathcal{J}^{x}_{\mathbf{k}}]_{\mu n,\nu m} [\mathcal{J}^{y}_{\mathbf{k}}]_{\nu m,\mu n} 
-[\mathcal{J}^{x}_{\mathbf{k}}]_{\mu n,\nu m} [\mathcal{J}^{y}_{\mathbf{k}}]_{\nu m,\mu n}}{(\epsilon_{\mu,n,\mathbf{k}} - \epsilon_{\nu,m,\mathbf{k}})^2} 
f_{\rm FD}(\epsilon_{\mu,n,\mathbf{k}} - \langle \mathcal{N} \rangle_{\mu,n,\mathbf{k}} \Omega),
\end{align}
%%%%%%%%%%%%%%%%%%%%%%%%%%%%%%%%%%%%%%%%%%%%%%%%%%%%%%%%%%%%%%%%%%%%%%%%%%
where the primed summation is carried over all possible values of $(\mu, n)$ and $(\nu, m)$ under the constraint that $(\mu,n)\neq(\nu,m)$.

Finally, we can derive the nonequilibrium Kubo formula for anomalous Hall conductivity in the impurity-free limit as follows:
%%%%%%%%%%%%%%%%%%%%%%%%%%%%%%%%%%%%%%%%%%%%%%%%%%%%%%%
\begin{align}
\sigma^{\rm free}_{xy} =  \int \frac{d^{2}\mathbf{k}}{(2\pi)^2} \sum_{\mu,n} \mathcal{B}_{\mu,n,\mathbf{k}} 
f_{\rm FD}\left(\epsilon_{\mu,n,\mathbf{k}} - \langle \mathcal{N} \rangle_{\mu,n,\mathbf{k}} \Omega\right),
\end{align}
%%%%%%%%%%%%%%%%%%%%%%%%%%%%%%%%%%%%%%%%%%%%%%%%%%%%%%%
where $\mathcal{B}_{\mu,n,{\bf k}}$ is the Berry curvature:
%%%%%%%%%%%%%%%%%%%%%%%%%%%%%%%%%%%%%%%%%%%%%%%%%%%%%%%%%%%%%%%%%%%%%%%%%%
\begin{align}
\mathcal{B}_{\mu,n,{\bf k}}=
-i \sum_{\substack{(\nu, m)\\(\nu, m)\neq(\mu, n)}}  \frac{[\mathcal{J}^{x}_{\mathbf{k}}]_{\mu n,\nu m} [\mathcal{J}^{y}_{\mathbf{k}}]_{\nu m,\mu n} 
-[\mathcal{J}^{x}_{\mathbf{k}}]_{\mu n,\nu m} [\mathcal{J}^{y}_{\mathbf{k}}]_{\nu m,\mu n}}{(\epsilon_{\mu,n,\mathbf{k}} - \epsilon_{\nu,m,\mathbf{k}})^2} ,
\label{eq:Berry}
\end{align}
%%%%%%%%%%%%%%%%%%%%%%%%%%%%%%%%%%%%%%%%%%%%%%%%%%%%%%%%%%%%%%%%%%%%%%%%%%
whose numerical values can be efficiently obtained using the Fukui-Hatsugai-Suzuki method~\cite{Fukui1}.

%%%%%%%%%%%%%%%%%%%%%%%%%%%%%%%%%%%%%%%%%%%%%%%%%%%%%%%
\section{Nonequilibrium Kubo formula for longitudinal conductivity in the fast-driving limit}
\label{appen:sigma_xx_fast}
%%%%%%%%%%%%%%%%%%%%%%%%%%%%%%%%%%%%%%%%%%%%%%%%%%%%%%%

We begin by expressing the nonequilibrium Kubo formula for longitudinal conductivity in a simplified notation:
%%%%%%%%%%%%%%%%%%%%%%%%%%%%%%%%%%%%%%%%%%%%%%%%%%%%%%%%%%%%%%%%%%%%%%%%%%%%
\begin{equation}
\sigma_{xx} =  \int \frac{d^{2}\mathbf{k}}{(2\pi)^2} \int \frac{d\omega}{2\pi} 
\mathrm{Tr} 
\left( 
J^x \frac{\partial G^{r}}{\partial \omega} J^x G^{<} -J^x G^{<} J^x \frac{\partial G^{a}}{\partial \omega}
\right),
\label{eq:sigma_xx_simplified}
\end{equation}
%%%%%%%%%%%%%%%%%%%%%%%%%%%%%%%%%%%%%%%%%%%%%%%%%%%%%%%%%%%%%%%%%%%%%%%%%%%%
where we have omitted the subscript, ${\bf k}$, of the current operator and the arguments, ${\bf k}$ and $\omega$, of the Green's functions for brevity.
Using the cyclic property of the ${\rm Tr}$ operator, we can further simplify Eq.~\eqref{eq:sigma_xx_simplified} as follows:
%%%%%%%%%%%%%%%%%%%%%%%%%%%%%%%%%%%%%%%%%%%%%%%%%%%%%%%%%%%%%%%%%%%%%%%%%%%%
\begin{equation}
\sigma_{xx} =  -\int \frac{d^{2}\mathbf{k}}{(2\pi)^2} \int \frac{d\omega}{2\pi} 
\mathrm{Tr} 
\left( 
J^x G^{<} J^x \frac{\partial (G^a-G^r)}{\partial \omega}
\right).
\label{eq:sigma_xx_simplified2}
\end{equation}
%%%%%%%%%%%%%%%%%%%%%%%%%%%%%%%%%%%%%%%%%%%%%%%%%%%%%%%%%%%%%%%%%%%%%%%%%%%%
The frequency integration in Eq.~\eqref{eq:sigma_xx_simplified2} can be transformed using integration by parts as follows:
%%%%%%%%%%%%%%%%%%%%%%%%%%%%%%%%%%%%%%%%%%%%%%%%%%%%%%%%%%%%%%%%%%%%%%%%%%%%
\begin{align}
\int \frac{d\omega}{2\pi} 
\mathrm{Tr} 
\left( 
J^x G^{<} J^x \frac{\partial (G^a-G^r)}{\partial \omega}
\right)
= 
&\frac{1}{2\pi}
\Big[
{\rm Tr} \left(J^x G^{<} J^x (G^a-G^r)\right)
\Big]^{\Omega/2}_{-\Omega/2}
\nonumber \\
&-\int \frac{d\omega}{2\pi} 
\mathrm{Tr} 
\left( 
J^x (G^a -G^r)J^x \frac{\partial G^<}{\partial \omega}
\right),
\end{align}
%%%%%%%%%%%%%%%%%%%%%%%%%%%%%%%%%%%%%%%%%%%%%%%%%%%%%%%%%%%%%%%%%%%%%%%%%%%%
where the first term on the right side of the equation becomes zero because of the Toeplitz property of Floquet matrices. 
Then, Equation~\eqref{eq:sigma_xx_simplified2} can be expressed as follows:
%%%%%%%%%%%%%%%%%%%%%%%%%%%%%%%%%%%%%%%%%%%%%%%%%%%%%%%%%%%%%%%%%%%%%%%%%%%%
\begin{equation}
\sigma_{xx} =  \int \frac{d^{2}\mathbf{k}}{(2\pi)^2} \int \frac{d\omega}{2\pi} 
\mathrm{Tr} 
\left( 
J^x (G^a-G^r) J^x \frac{\partial G^<}{\partial \omega}
\right).
\label{eq:sigma_xx_simplified3}
\end{equation}
%%%%%%%%%%%%%%%%%%%%%%%%%%%%%%%%%%%%%%%%%%%%%%%%%%%%%%%%%%%%%%%%%%%%%%%%%%%%

Now, in the fast-driving limit, i.e., $\Omega \rightarrow \infty$, we observe that the following property holds, with the dependencies on ${\bf k}$ and $\omega$ explicitly included: 
%%%%%%%%%%%%%%%%%%%%%%%%%%%%%%%%%%%%%%%%%%%%%%%%%%%%%%%%%%%%%%%%%%%%%%%%%%%%
\begin{align}
[\hat{J}^x_{\bf k}]_{nm} &\simeq [\hat{J}^{x}_{\bf k}]_{nn}\delta_{nm},
\label{eq:fast_limit1} \\
[\hat{G}^{r}({\bf k},\omega)]_{nm} &\simeq [\hat{G}^{r}({\bf k},\omega)]_{nn}\delta_{nm},
\label{eq:fast_limit2} \\
[\hat{G}^<({\bf k},\omega)]_{nm} &\simeq \left[\hat{G}^a({\bf k},\omega)-\hat{G}^r({\bf k},\omega)\right]_{nn} f_{{\rm loc},n}(\omega) \delta_{nm},
\label{eq:fast_limit3}
\end{align}
%%%%%%%%%%%%%%%%%%%%%%%%%%%%%%%%%%%%%%%%%%%%%%%%%%%%%%%%%%%%%%%%%%%%%%%%%%%%
where $[\hat{J}^x_{\bf k}]_{nm}$ represents the $(n,m)$-th Floquet matrix component of the current operator, expressed as a $2 \times 2$ matrix with respect to the sublattice index. 
Also, $[\hat{G}^{a,r,<}({\bf k},\omega)]_{nm}$ represents the $(n,m)$-th Floquet matrix component of the advanced, retarded, and lesser Green's functions, respectively, expressed as a $2 \times 2$ matrix with respect to the sublattice index. 
Importantly, $f_{{\rm loc},n}(\omega)$ is independent of the sublattice index and momentum.
Equations~\eqref{eq:fast_limit1}, \eqref{eq:fast_limit2}, and \eqref{eq:fast_limit3} hold in the fast-driving limit because all the Floquet eigenstates are well separated from each other by a significant quasienergy gap on the order of $\Omega$.

Inserting Eqs.~\eqref{eq:fast_limit1}, \eqref{eq:fast_limit2}, and \eqref{eq:fast_limit3} into Eq.~\eqref{eq:sigma_xx_simplified3}, we can obtain the nonequilibrium Kubo formula for longitudinal conductivity in the fast-driving limit as follows:
%%%%%%%%%%%%%%%%%%%%%%%%%%%%%%%%%%%%%%%%%%%%%%%%%%%%%%%%%%%%%%%%%%%%%%%%%%%%
\begin{align}
\sigma_{xx}^{\rm fast} =  \int \frac{d^{2}\mathbf{k}}{(2\pi)^2} \int \frac{d\omega}{2\pi} 
\sum_n \mathrm{Tr}^\prime 
\Bigg( 
&\left[\hat{J}^x (\hat{G}^a-\hat{G}^r) \hat{J}^x \frac{\partial (\hat{G}^a-\hat{G}^r)}{\partial \omega}\right]_{nn}
f_{{\rm loc},n}(\omega)
\nonumber \\
+&
\left[\hat{J}^x (\hat{G}^a-\hat{G}^r) \hat{J}^x (\hat{G}^a-\hat{G}^r)\right]_{nn}
\frac{\partial f_{{\rm loc},n}(\omega)}{\partial \omega}
\Bigg).
\label{eq:sigma_xx_fast1}
\end{align}
%%%%%%%%%%%%%%%%%%%%%%%%%%%%%%%%%%%%%%%%%%%%%%%%%%%%%%%%%%%%%%%%%%%%%%%%%%%%
where ${\rm Tr}^\prime$ indicates that the trace is taken only over the sublattice indices.  

On the other hand, if we insert Eqs.~\eqref{eq:fast_limit1}, \eqref{eq:fast_limit2}, and \eqref{eq:fast_limit3} into Eq.~\eqref{eq:sigma_xx_simplified2}, we obtain an alternative expression for the nonequilibrium Kubo formula for longitudinal conductivity in the fast-driving limit as follows:
%%%%%%%%%%%%%%%%%%%%%%%%%%%%%%%%%%%%%%%%%%%%%%%%%%%%%%%%%%%%%%%%%%%%%%%%%%%%
\begin{align}
\sigma_{xx}^{\rm fast} =  -\int \frac{d^{2}\mathbf{k}}{(2\pi)^2} \int \frac{d\omega}{2\pi} 
\sum_n \mathrm{Tr}^\prime 
\left[\hat{J}^x (\hat{G}^a-\hat{G}^r) \hat{J}^x \frac{\partial (\hat{G}^a-\hat{G}^r)}{\partial \omega}\right]_{nn}
f_{{\rm loc},n}(\omega),
\label{eq:sigma_xx_fast2}
\end{align}
%%%%%%%%%%%%%%%%%%%%%%%%%%%%%%%%%%%%%%%%%%%%%%%%%%%%%%%%%%%%%%%%%%%%%%%%%%%%
which is identical to the first term on the right side of Eq.~\eqref{eq:sigma_xx_fast1}, but includes a minus sign.
Combining Eq. \eqref{eq:sigma_xx_fast1} and Eq. \eqref{eq:sigma_xx_fast2}, we arrive at the following equation:
%%%%%%%%%%%%%%%%%%%%%%%%%%%%%%%%%%%%%%%%%%%%%%%
\begin{align}
\sigma_{xx}^{\rm fast} =  \frac{1}{2} \int \frac{d^{2}\mathbf{k}}{(2\pi)^2} \int \frac{d\omega}{2\pi} 
\sum_n \mathrm{Tr}^\prime 
\left[\hat{J}^x (\hat{G}^a-\hat{G}^r) \hat{J}^x (\hat{G}^a-\hat{G}^r)\right]_{nn}
\frac{\partial f_{{\rm loc},n}(\omega)}{\partial \omega}.
\label{eq:sigma_xx_fast3}
\end{align}
%%%%%%%%%%%%%%%%%%%%%%%%%%%%%%%%%%%%%%%%%%%%%%%

Now, using the definition of the spectral weight, $\rho_{\bf k}(\omega)$,
%%%%%%%%%%%%%%%%%%%%%%%%%%%%%%%%%%%%%%%%%%%%%%%
\begin{align}
\hat{G}^a({\bf k},\omega)-\hat{G}^r({\bf k},\omega)=  2\pi i \rho_{\bf k}(\omega),
\label{eq:rho_Floquet}
\end{align}
%%%%%%%%%%%%%%%%%%%%%%%%%%%%%%%%%%%%%%%%%%%%%%%
we can rewrite Eq.~\eqref{eq:sigma_xx_fast3} as follows:
%%%%%%%%%%%%%%%%%%%%%%%%%%%%%%%%%%%%%%%%%%%%%%%
\begin{align}
\sigma_{xx}^{\rm fast} =  2\pi^2 \int \frac{d^{2}\mathbf{k}}{(2\pi)^2} \int \frac{d\omega}{2\pi} 
\sum_n \mathrm{Tr}^\prime 
\left[\hat{J}^x_{\bf k} \rho_{\bf k}(\omega) \hat{J}^x_{\bf k} \rho_{\bf k}({\omega}) \right]_{nn}
\left(-\frac{\partial f_{{\rm loc},n}(\omega)}{\partial \omega}\right).
\label{eq:sigma_xx_fast4}
\end{align}
%%%%%%%%%%%%%%%%%%%%%%%%%%%%%%%%%%%%%%%%%%%%%%%

The final task is to determine $f_{{\rm loc},n}(\omega)$ so that it remains valid not only in the fast-driving limit but also potentially beyond that.
To this end, we define $f_{{\rm loc},n}(\omega)$ by taking the average over momenta and the trace over sublattice indices on both sides of Eq.~\eqref{eq:fast_limit3}, which leads to
%%%%%%%%%%%%%%%%%%%%%%%%%%%%%%%%%%%%%%%%%%%%%%%%%%%%%%%%%%%%%%%%%%%%%%%%%%%%
\begin{align}
f_{{\rm loc},n}(\omega) = f_{\rm loc}(\omega+n\Omega) =
\frac{\int \frac{d^{2}\mathbf{k}}{(2\pi)^2}{\rm Tr}^\prime[\hat{G}^<({\bf k},\omega)]_{nn}}
{\int \frac{d^{2}\mathbf{k}}{(2\pi)^2}{\rm Tr}^\prime \left[\hat{G}^a({\bf k},\omega)-\hat{G}^r({\bf k},\omega)\right]_{nn}},
\end{align}
%%%%%%%%%%%%%%%%%%%%%%%%%%%%%%%%%%%%%%%%%%%%%%%%%%%%%%%%%%%%%%%%%%%%%%%%%%%%
which is equal to the local distribution function, $f_{\rm loc}(\omega) = N_{\rm loc}(\omega)/\rho_{\rm loc}(\omega)$, defined in the main text.

%%%%%%%%%%%%%%%%%%%%%%%%%%%%
\section{Robustness of negative resistance catastrophe}
\label{appen:disorder_effect}
%%%%%%%%%%%%%%%%%%%%%%%%%%%%

%%%%%%%%%%%%%%%%%%%%%%%%%%%%%%%%%%%%%%%%%%%%%%%%%%%%%%%%%%%%%%%%%%%%%
\begin{figure}[h!]
\includegraphics[width=0.8\linewidth]{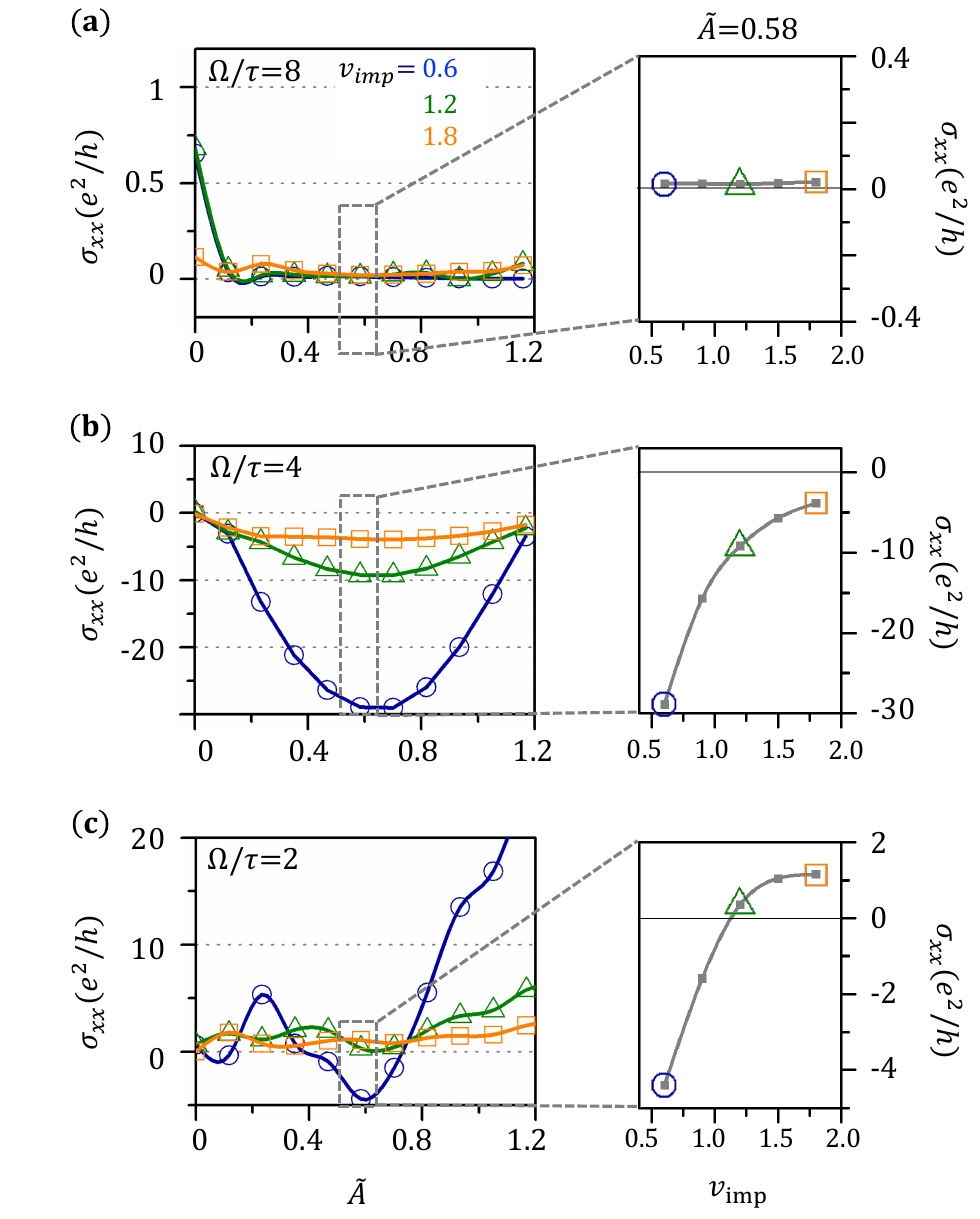}
\caption{\textbf{Stability of negative longitudinal conductivity against impurity scattering.}
The nonequilibrium longitudinal conductivity calculated from the fully interacting Green's functions, $\sigma_{xx}^{\rm full}$, is computed as a function of $\tilde{A}$ at ({\bf a}) $\Omega/\tau=8$, ({\bf b}) $\Omega/\tau=4$, and ({\bf c}) $\Omega/\tau=2$ for different values of impurity scattering strength, $v_{\rm imp}$.
The panels on the right, which are zoomed-in figures of the dashed boxes, show in detail how $\sigma^{\rm full}_{xx}$ at $\tilde{A}=0.58$ varies with $v_{\rm imp}$. 
}
\label{FIG6}
\end{figure}
%%%%%%%%%%%%%%%%%%%%%%%%%%%%%%%%%%%%%%%%%%%%%%%%%%%%%%%%%%%%%%%%%%%%%

In this section, we assess the robustness of negative longitudinal conductivity and, consequently, NRC by varying the impurity scattering strength.
To this end, let us review how negative longitudinal conductivity will trigger NRC and thus cause zero resistance in irradiated graphene, based on the same mechanism developed for the quantum RIZR state~\cite{Andreev1}.

When negative longitudinal conductivity occurs, the current-voltage characteristic curve adopts an S-shaped form.
The S-shaped current-voltage characteristic curve requires that two specific finite currents with opposite signs be present at zero electric field. 
This triggers a catastrophic breakdown where the current distribution spontaneously becomes rotational, meaning that $\nabla\times{\bf J} \neq 0$, even when the electric field is zero. 
This embodies the core of NRC.
Meanwhile, the net electric field inside the system (with no contributions from irradiation) is irrotational since $\nabla\times{\bf E}=0$.
If the net electric field is related to the current through Ohm's law, these two conditions cannot be satisfied simultaneously.
Therefore, the longitudinal conductivity must go to zero so that the net electric field and the current become decoupled, at least for the dissipative channel.   
It is worth noting that the anomalous Hall conductivity remains finite in irradiated graphene, similar to the quantum Hall RIZR state.
With a finite anomalous Hall conductivity, inverting the conductivity matrix reveals that both longitudinal conductivity and resistivity are zero.
Note that an alternative mechanism has also been proposed to explain the quantum Hall RIZR state~\cite{Inarrea1,Inarrea2}.

Now, let us investigate how robust NRC can be against impurity scattering, which is expected to favor a more traditional Drude-like behavior.
To this end, we analyze how the nonequilibrium longitudinal conductivity changes with various impurity scattering strengths.
Figure~\ref{FIG6} plots $\sigma_{xx}^{\rm full}$ as a function of $\tilde{A}$ at ({\bf a}) $\Omega/\tau=8$, ({\bf b}) $\Omega/\tau=4$, and ({\bf c}) $\Omega/\tau=2$ for different values of impurity scattering strength, $v_{\rm imp}$.
Specifically, Figure~\ref{FIG6}~({\bf a}) shows that, at $\Omega/\tau=8$, $\sigma^{\rm full}_{xx}$ begins with a positive value at $\tilde{A}=0$, which, corresponding to the longitudinal conductivity of graphene in equilibrium, aligns with the well-known universal conductivity of $2/\pi$ in units of $e^2/h$~\cite{Shon1} until impurity scattering becomes too strong.
Note that $\sigma^{\rm full}_{xx}$ decreases from this universal conductivity at sufficiently strong impurity scattering, e.g., $v_{\rm imp}=1.8$.
As $\tilde{A}$ increases, the Floquet Chern insulator emerges, as indicated by nearly zero $\sigma^{\rm full}_{xx}$, which remains stable against impurity scattering.
Meanwhile, Figure~\ref{FIG6}~({\bf b}) shows that, at $\Omega/\tau=4$, $\sigma^{\rm full}_{xx}$ is negative over a wide range of $\tilde{A}$, but its magnitude decreases as impurity scattering becomes stronger.
However, the sign of $\sigma^{\rm full}_{xx}$ remains negative even until $v_{\rm imp}=1.8$. 
By contrast, Figure~\ref{FIG6}~({\bf c}) shows that, at $\Omega/\tau=2$, $\sigma^{\rm full}_{xx}$ changes its sign from negative to positive around $v_{\rm imp}=1.0$. 
This suggests that NRC can be robust against impurity scattering at $\Omega \simeq {\cal W}/2$, where negative longitudinal conductivity is strongest.
However, NRC can be suppressed at lower $\Omega/\tau$ with sufficiently strong impurity scattering.

% Produces the bibliography via BibTeX.
\bibliography{FCI_RIZR_in_irradiated_graphene}

\end{document}